\documentclass[%
 reprint,
superscriptaddress,
 amsmath,amssymb,
 aps,
pra,
]{revtex4-2}

\usepackage{tikz,xcolor}
\usepackage{graphicx}
\usepackage{dcolumn}
\usepackage{bm}
\usepackage{hyperref}
\usepackage{subcaption}
\usepackage{dsfont}
\usepackage{dutchcal}
\usepackage{mathrsfs}
\usepackage{verbatim}
\usepackage{float}
\usepackage[title]{appendix}
\usepackage{physics}
\hypersetup{
    colorlinks=true,
    urlcolor= blue,
    citecolor=blue,
    linkcolor= blue}
\def\d{\mbox{\rm d}}
\def\e{\mbox{\rm e}}

\definecolor{lime}{HTML}{A6CE39}
\DeclareRobustCommand{\orcidicon}{%
	\begin{tikzpicture}
	\draw[lime, fill=lime] (0,0) 
	circle [radius=0.16] 
	node[white] {{\fontfamily{qag}\selectfont \tiny ID}};
	\draw[white, fill=white] (-0.0625,0.095) 
	circle [radius=0.007];
	\end{tikzpicture}
	\hspace{-2mm}
}

\foreach \x in {A, ..., Z}{%
	\expandafter\xdef\csname orcid\x\endcsname{\noexpand\href{https://orcid.org/\csname orcidauthor\x\endcsname}{\noexpand\orcidicon}}
}

\begin{document}


\title{Quantum regression theorem in the Unruh–DeWitt battery}

\author{Manjari Dutta\orcidA{}}
\email{chandromouli15@gmail.com}

\author{Arnab Mukherjee\orcidB{}}
\email{mukherji.arn@gmail.com}

\author{Sunandan Gangopadhyay\orcidC{}}
\email{sunandan.gangopadhyay@gmail.com}
\affiliation{Department of Astrophysics and High Energy Physics,\linebreak
	S.N.~Bose National Centre for Basic Sciences,\linebreak
	 JD Block, Sector-III, Salt Lake, Kolkata 700106, India}

\begin{abstract}
\noindent In this paper, we employ the \textit{quantum regression theorem}, a powerful tool in the study of open quantum systems, to analytically study the correlation functions of an \textit{Unruh–DeWitt detector}, which is an uniformly accelerated two-level quantum system, absorbing charges from an external classical coherent pulse. The system can thus be viewed as a relativistic quantum battery that interacts with the environment of its perceived particles, namely, the quanta of a massless scalar field. By considering the relativistic battery moving in \textit{Rindler spacetime}, under \textit{Born-Markov approximation}, we derive the \textit{Gorini-Kossakowski-Sudarshan-Lindblad master equation} governing the evolution of the system's reduced density matrix. Moreover, we perform the Fourier transformation of the \textit{Wightman functions} and use \textit{exponential regularisation} to compute the functional forms appearing in the master equation. Next, we derive the evolution equations for the the single-time expectation values of the system's operators. We not only solve these equations to find out the single time averages, but also employ the quantum regression theorem to determine the two-time correlation functions of first and second order. We analyse them in details to explain the phenomenon of \textit{spontaneous emission} and show analytically how the acceleration can enhance the associated dissipation. Furthermore, we address a special form of second order correlation function relevant to the context of \textit{photon bunching} arising in Bose-Einstein statistics. We analysed the results both analytically and graphically. Finally, we derive the spontaneous emission spectrum of the battery detector analytically, which in the long-time limit displays a well-defined \textit{Lorentzian line shape} in the high frequency regime.  
  

\end{abstract}

\maketitle

\section{Introduction}\label{S:1}
\noindent Over the last two decades due to the rapid advances in quantum technologies, quantum thermodynamics emerged as a very active field of research \cite{kosloff2013quantum, vinjanampathy2016quantum, bhattacharjee2021quantum, campaioli2024colloquium}. The primary concern of this field is to study the laws of thermodynamics and quantities like work, heat, and energy in the length scales of the order of nano-meters \cite{cengel2011thermodynamics}. With continued technological miniaturization, precise control and management of energy at microscopic scales becomes essential for the reliable and efficient operation of modern devices. This leads to the upsurge of interest in the theoretical and experimental study of an alternative and efficient energy storage device, known as a quantum battery \cite{alicki2013entanglement, campaioli2018quantum}. 

Using the non-classical properties of quantum systems, such as quantum superposition, coherence, entanglement, and many-body collective behaviours, quantum batteries are able to perform faster and more efficient charging processes than their classical counterparts \cite{campisi2011colloquium, horodecki2013fundamental, goold2016role, campaioli2018quantum, binder2015quantacell, campaioli2017enhancing, farina2019charger, zhang2019powerful, carrasco2022collective, gyhm2022quantum, rodriguez2023catalysis, gemme2023offresonant, gemme2024qutrit, song2024remote}. Quantum batteries can be realised through various systems such as fluorescent organic molecules embedded in a microcavity \cite{quach2022superabsorption}, IBM quantum chips \cite{gemme2022ibm}, transmon qubits \cite{hu2022optimal}, quantum dots \cite{wenniger2023experimental}, two-level system coupled to a photonic waveguide \cite{lu2025topological, tirone2025manybody}, based on the Sachdev-Ye-Kitaev (SYK) model \cite{rossini2020quantum, rosa2020ultra, kim2022operator}, and based on the collision model \cite{shaghaghi2022micromaser, rodriguez2023artificial}. The interaction of a quantum battery with its surroundings results in the coherence of the battery leaking to the surroundings. Due to this decoherence effect, the charging and discharging performance of quantum batteries is frequently affected \cite{farina2019charger, qb_prl, carrega2020dissipative, tabesh2020environment}.

The study of quantum batteries in the relativistic framework is inspired from the fundamental perspective of understanding quantum effects in the
relativistic regime, and no less from the current technological outlook of satellite based quantum networks such as the quantum internet \cite{lu2022micius, ribezzo2023deploying}. In recent times, it has been noticed that studying quantum heat engines in the relativistic framework leads to some exciting results in the context of work extraction and efficiency \cite{arias2018unruh, mukherjee2022unruh}. Following this line of work, the charging process and the work extraction of a quantum battery have been studied in a relativistic framework \cite{hao2023quantum, tian2025dissipative, mukherjee2024enhancement, chen2025quantum}. Much of the existing work on quantum batteries in this context adopts the minimal quantum-detector paradigm,
namely, a two-level system (qubit), widely known as the Unruh--DeWitt detector \cite{dewitt1979general, unruh1984what}. In this framework, as the detector always interacts with the surrounding quantum fields, this framework is known as the \textit{open quantum framework}. In the open quantum framework, a plethora of works have been investigated using a single \cite{benatti2004entanglement, yu2008understanding, hu2012geometric, jin2014dynamical,liu2021relativistic} and two Unruh-DeWitt detectors \cite{benatti2004entanglement,zhang2007unruh,hu2015entanglement,yang2016entanglement,cheng2018entanglement,she2019entanglement,zhou2021entanglement, chen2022entanglement} in diverse physical scenarios.

In the context of the open quantum framework, the dynamics of the model is usually studied by two types of master equations. These are the Markovian master equations, known as the \textit{Gorini-Kossakowski-Sudarshan-Lindblad (GKSL)} quantum master equation \cite{gorini1976completely, lindblad1976generators} and the non-Markovian quantum master equations, such as 
the \textit{Nakajima-Zwanzig (NZ)} equation \cite{nakajima1958general, zwanzig1960ensemble} and the \textit{time-convolutionless (TCL)} equation \cite{shibata1977generalized, chaturvedi1979time, shibata1980expansion}. The form of the GKSL master equation for a model like the \textit{Unruh-DeWitt (UDW) battery} under a coherent drive with linear and quadratic detector-field coupling has been studied in \cite{mukherjee2024enhancement}. These master equations evaluate the reduced density matrix of the system at a later time. Apart from using the master equation technique directly, the reduced density matrix can be computed by the \textit{Feynman-Vernon influence functional method} \cite{feynman1963theory}. The Feynman–Vernon influence-functional formalism provides a systematic route to the reduced dynamics of an open quantum system by integrating out the environmental degrees of freedom. The resulting reduced density matrix captures the system’s evolution under environmental back-action without requiring explicit access to the environment. Moreover, the same framework enables the evaluation of relevant correlation functions, including system–environment coherences and energy-transfer characteristics from the system into the environment. The influence-functional formalism is very effective in both the Markovian as well as non-Markovian scenario. Although quantum master equation approaches and the influence-functional formalism provide convenient routes to the system’s reduced density matrix, they are generally not well suited for computing multi-point correlation functions. In their standard implementations, both frameworks are primarily built around two-point environmental correlators, making higher-order (multi-time) correlations difficult to access without additional approximations or extensions. At this point \textit{quantum regression theorem (QRT)} plays an important role \cite{Breuer2007, carmichael1999statistical}.\\[3pt]
\noindent The term \textit{regression} literally refers to a relationship. Generally, regression is a statistical and machine learning technique that is used to model and analyse the relationship between the input variables which are provided independently and the target variables which are to be predicted by identifying the pattern in input variables. In the context of open quantum systems, QRT is a very powerful tool to compute the multi-time correlation functions using similar dynamics governed by the single time expectation values. This is because, unlike the single time average, a multi-time average value cannot be obtained from the master equation directly. Hence, QRT is a derived relation that can correctly predict the dynamics exhibited by multi-time correlation functions, following the same governing pattern as single-time averages. There are several physical phenomena, which are experimentally observable, and are linked intimately with  
the correlation functions of the system, such as power spectrum \cite{carmichael1999statistical, rf-spectrum1, rf-spectrum2} of spontaneous emission, absorption and resonance fluorescence in the field of spectroscopy as well as \textit{photon bunching or anti bunching phenomena} \cite{antibunch1, Phonon-photon, antibunch2} associated with \textit{Hanbury Brown and Twiss (HBT)} effect in the field of quantum optics. \\[3pt]
\noindent In the literature, there are many studies \cite{qrt-fluc, qrt-noise, qrt_laser-master} where QRT has been successfully and significantly used to compute the two-time correlators. It was more widely used to develop the \textit{theory of resonance fluorescence} (\cite{qrt-res-flur1, qrt-res-flur2, qrt-res-flur3}). Later, this theorem was derived and the corresponding correlators were obtained by using the master equation approach \cite{qrt-master1, qrt-master2}. In most studies in the literature, the usefulness and suitability of QRT for calculating correlation functions are conventionally restricted to the weakly coupled regime and Markovian dynamics, where the correlation functions obtained from QRT are simply two-time averages of the system’s operators. Numerous studies \cite{qrt-qdot, non-mark1, qrt-non-mark, non-mark2, non-mark3, non-mark4, non-mark5, Rivas_Hulega_non_mark} in the literature discuss the failure of QRT in the context of non-Markovian dynamics. However, to address such kind of limitations, efforts have also been made to modify the standard conventional method of QRT. In \cite{qrt-otoc}, for the system following Markovian dynamics, an extended version of QRT is introduced to compute \textit{out-of-time-order correlation} functions. Here, we would like to refer to another interesting in \cite{qrt-hei}, where QRT is derived by using the Heisenberg 
picture \cite{oqs-hei} to obtain the multi-time correlators for a system undergoing Markovian dynamics. By using this methodology in their subsequent work, \cite{qrt_Kubo-Martin-Schwinger}, they introduced a modification of the QRT under Markov approximation. Furthermore, they also showed that the two-point correlators produced by their modified form of QRT satisfy the \textit{Kubo–Martin–Schwinger equilibrium condition} up to the non-zero order in the system–environment coupling, whereas the conventional form of QRT cannot perform the required task. It is also relevant to mention their most recent study \cite{qrt-ultrastrong}, in which two-time correlators are derived in the \textit{ultrastrong-coupling (USC)} regime between the system and the environment. \\[3pt]
\noindent In our present study, although we adopt the standard conventional form of QRT which is applicable only for obtaining the two-time correlation functions for a weakly-coupled system undergoing Markov dynamics, we assign the corresponding system to be a relativistic Unruh-Dewitt battery which is a less explored system in the context of correlation functions and the associated physical phenomena. As also argued in \cite{qrt-ultrastrong}, the correlators, in comparison to the reduced density matrix, can frequently reveal deeper insights \cite{carmichael1999statistical, corre-insight1, corre-insight2, corre-insight3} into both the dynamical and steady-state behaviour of a system. It motivates us to analyse a Unruh-DeWitt battery from this perspective. Moreover, since our battery accelerates with respect to an inertial frame, the appearance of the environmental bath and the openness of the system as a quantum system depends on its uniform acceleration. Therefore, we also intend to investigate how this acceleration influences the correlation functions and the resulting physical features of interest.  \\[2pt]
\noindent The work carried out in this paper are in natural units ($\hbar=1, c=1$) and it is organized as follows. In section \ref{S:2}, we provide a brief review on the concept of quantum regression theorem, figuring out the main key steps in its implementation, and presenting examples of evolution equations for the first and second order correlation functions required for our present study of a weakly coupled Markovian system. In section \ref{S:3}, we consider a time independent Hamiltonian form of a two level quantum system and couple it linearly to an external classical time dependent pulse. We then construct the system Hamiltonian of the Unruh-DeWitt battery. We also introduce the Hamiltonian of the bath of massless scalar field and present the interaction Hamiltonian as a function of the proper time in the interaction picture. In section \ref{S:4}, we consider the weak-coupling between the system and the bath and derive the evolution equation of the reduced density matrix of the system in the Schr\"{o}dinger picture. We apply Rindler mapping, parametrize the coordinates of the uniformly accelerating detector in terms of its proper time, and calculate the Wightman function. We then perform the Fourier transform of the Wightman function and determine its real and imaginary part in the dual space of proper time. We also use exponential regularization to extract a finite contribution from the divergent imaginary part. In section \ref{S:5}, we define a set of ladder operators for the system Hamiltonian. Next, we calculate the matrix elements of the master equation and determine the evolution of the single time expectation values of the Pauli operators. From these results, we obtain the evolution equations governing the ladder and number operators and solve them to find their single time averages. We also define the fluctuations of those operators with respect to their steady state and present the evolution of the single time averages of those fluctuations. In next section \ref{S:6}, we directly apply QRT and calculate the corresponding fluctuation correlation functions. From these, we compute the first and second order correlators and analyse them to explain various physical phenomena such as spontaneous emission and Hanbury-Brown-Twiss effect. Next, In next section \ref{S:7}, we obtain the power spectrum of the spontaneous emission and analyse the result analytically. Finally, in section \ref{S:8}, we conclude our study.  
\section{Quantum Regression Theorem: Basic Overview}\label{S:2}
\noindent In this section, we give a brief overview of the \textit{quantum regression theorem (QRT)} \cite{carmichael1999statistical, Breuer2007}. As we have discussed in the earlier section, the standard and conventional form of QRT implies that the dynamics of the evolution of the single time average determines the same of the two-time average. Generally, the evolution of the single-time average of any operator $\mathcal{\hat{O}}$ can be found from the relation \cite{carmichael1999statistical, Breuer2007}
\begin{align}
\big<\dot{\mathcal{\hat{\,O}}}\big>\,=\,Tr\big[\mathcal{\hat{\,O}}\,\dot{\rho}(t)\big]~,   
\end{align}
where $\rho(\tau)$ is the reduced density matrix of the system of interest and the evolution of a multi-time correlation function cannot be presented in this manner. One has to therefore proceed in the following way.\\  [2pt]
\noindent We begin by considering $\hat{\mathcal{A}}_{\mu}$ \big($\mu=1,2,...,$\big), a complete set of operators acting on the Hilbert space of the system such that the time-dependent expectation value of the operator reads 
\begin{align}
\big<\dot{\hat{\mathcal{A}}}_{\mu}\big>\,=\,\sum_{\lambda}M_{\mu\lambda}\big<\hat{\mathcal{A}}_{\mu}\big>~.\label{e.2.1}
\end{align}
Then the aforementioned coupled set of linear differential equations can be presented in the following vector form
\begin{align}
\big<\dot{\boldsymbol{\hat{\mathcal{A}}}}\big>\,=\,\boldsymbol{M}\,\big<\boldsymbol{\hat{\mathcal{A}}}\big>~;\label{e.2.2}
\end{align}
where $\hat{\boldsymbol{\mathcal{A}}}$ is a column vector of operators and $M$ is an evolution matrix.\\
\noindent Now, regarding the computation of the two-time correlation functions, the key results provided by the quantum regression theorem, for ($\tau\,\geq0$) reads
\begin{align}
&\dfrac{d}{d\tau}\Big<\hat{\mathcal{O}}(t)\,\hat{\mathcal{A}}_{\mu}(t+\tau)\Big>\,=\,\sum_{\lambda}M_{\mu\lambda}\,\Big<\hat{\mathcal{O}}(t)\,\hat{\mathcal{A}}_{\mu}(t+\tau)\Big>       ,         \nonumber\\
\Rightarrow\,&\dfrac{d}{d\tau}\Big<\hat{\mathcal{O}}(t)\,\hat{\boldsymbol{\mathcal{A}}}(t+\tau)\Big>\,=\,\boldsymbol{M}\,\Big<\hat{\mathcal{O}}(t)\,\hat{\boldsymbol{\mathcal{A}}}(t+\tau)\Big>~;\label{e.2.3}  
\end{align}
where $\hat{\mathcal{O}}$ is allowed to be any operator acting on the Hilbert space of the system, including the components of $\hat{\mathcal{A}}_{\mu}$. To explain in a more clear manner, corresponding to each operator $\hat{\mathcal{O}}$, there exists a set of correlation functions $\big<\hat{\mathcal{O}}(t)\,\hat{\mathcal{A}}_{\mu}(t+\tau)\big>$, \big($\mu\,=\,1, 2, .....$\big), which govern the same differential equations (with respect to $\tau$), as $\mathcal{A}_{\mu}(t+\tau)$, the set of single time averages does.\\ 
\noindent The two time correlation functions generated by multiplying a single operator with $\hat{\boldsymbol{\mathcal{A}}}$ is known as first order correlation function. However, this order can be chosen to be higher. For example, by multiplying $\hat{\mathcal{O}}^{\prime}(t)$, another operator to the right of $\hat{\mathcal{O}}(t)\,\hat{\boldsymbol{\mathcal{A}}}(t+\tau)$, a two time correlation function of second order can be obtained and it governs
\begin{align}
\dfrac{d}{d\tau}\Big<\hat{\mathcal{O}}(t)\,\hat{\mathcal{A}}_{\mu}(t+\tau)&\,\hat{\mathcal{O}}^{\prime}(t)\Big>\nonumber\\
=&\,\sum_{\lambda}M_{\mu\lambda}\,\Big<\hat{\mathcal{O}}(t)\,\hat{\mathcal{A}}_{\mu}(t+\tau)\,\hat{\mathcal{O}}^{\prime}(t)\Big>,\nonumber\\
\Rightarrow\,\dfrac{d}{d\tau}\Big<\hat{\mathcal{O}}(t)\,\hat{\boldsymbol{\mathcal{A}}}(t+\tau)\,&\hat{\mathcal{O}}^{\prime}(t)\Big>\,=\,\boldsymbol{M}\,\Big<\hat{\mathcal{O}}(t)\,\hat{\boldsymbol{\mathcal{A}}}(t+\tau)\,\hat{\mathcal{O}}^{\prime}(t)\Big>~.\label{e.2.4}  
\end{align}
Now, by solving the above eqs. (\ref{e.2.3}, \ref{e.2.4}), the correlators can easily be obtained.\\
\noindent The aforementioned version of QRT is sufficient for the present study, as the system is weakly coupled, Markovian (memoryless), and initially uncorrelated. As we have already discussed in the earlier section, this theory may become invalid and require modifications in the presence of initial system–bath correlations, strong system–bath coupling, or non-Markovian system dynamics.

\section{Model of Unruh-DeWitt Battery}\label{S:3}
\noindent In this section, we recapitulate a concise formulation of the model describing a relativistic quantum battery driven by an external field, modelled through a single UDW detector interacting with a quantum field \cite{mukherjee2024enhancement}.

We model the UDW detector as a two-level quantum system with states $\ket{g}$ and $\ket{e}$, corresponding to the ground and excited energy levels, respectively. The associated energy eigenvalues are $-\omega_{0}/2$ for $\ket{g}$ and $+\omega_{0}/2$ for $\ket{e}$. Initially, the detector is assumed to be prepared in its ground state $\ket{g}$. The free Hamiltonian of the detector is therefore expressed as (with $\hbar =1$)
\begin{equation}
    H_{D}=\frac{\omega_{0}}{2}\sigma_z~,\label{e_3.1}
\end{equation}
where $\sigma_z = \ket{e}\bra{e} - \ket{g}\bra{g}$ denotes the Pauli–Z operator. Upon applying a time-dependent classical driving pulse to the detector, the system behaves as a driven two-level quantum battery, and the corresponding total Hamiltonian takes the form \cite{gemme2022ibm}
\begin{equation}
H'_{D}=\frac{\omega_{0}}{2}\sigma_z+\Omega f(t) (\sigma_{+}+\sigma_{-})~,\label{e_3.2}
\end{equation}
where $f(t)=f_0\cos{(\omega t)}$ is the time dependent amplitude of the classical pulse, and $\mu$ is the effective coupling strength between the UDW detector and the classical drive. Here we set $f_0=1$ for $0< t\leq \tau$, where $\tau$ is the total charging time \cite{hao2023quantum}.
Since the intrinsic frequency of the detector satisfies $\omega_{0} \gg \Omega$, efficient charging of the quantum battery requires the external driving field to be resonant with the detector’s energy gap $\omega_{0}$~\cite{carrega2020dissipative}, that is, $\omega = \omega_{0}$, known as the \textit{resonance condition}. Under this condition and in the weak-coupling limit, the total Hamiltonian of the quantum battery, within the rotating-wave approximation (RWA), takes the form~\cite{scully1997quantum}
\begin{equation}
H''_{D}=\frac{\omega_{0}}{2}\sigma_z+\frac{\Omega}{2} (\sigma_{+}e^{-i\omega_{0} \tau}+\sigma_{-}e^{i\omega_{0} \tau})~.\label{e_3.3}
\end{equation}\\[2pt]
To get rid of this time dependency and obtain a simpler form of the Hamiltonian, we take a rotating frame transformation \cite{gemme2022ibm}
\begin{equation}
    H_{B}=UH''_{D}U^{\dagger}-iU\frac{d}{d\tau}U^{\dagger}~,\label{e_3.4}
\end{equation}
where $U$ is the unitary operator. Taking $U=e^{iH_D \tau}=e^{i\frac{\omega_{0}}{2}\sigma_z \tau}$, and putting eq.~\eqref{e_3.3} in eq.~\eqref{e_3.4}, the total Hamiltonian of the quantum battery becomes
\begin{equation}
H_{B}=\frac{\Omega}{2}(\sigma_{+} +\sigma_{-})\,=\,\Omega\,\sigma_x~,\label{e_3.5}
\end{equation}
where $\sigma_{+}=\vert e\rangle\langle g\vert$ and $\sigma_{-}=\vert g\rangle\langle e\vert$ are the raising and lowering operators.

In the following section, we aim to investigate the influence of the environment on the performance of the quantum battery by analysing its interaction with a massless quantum scalar field which has the following Hamiltonian,
\begin{equation}
    H_{\phi}=\int_{-\infty}^{+\infty}\,d^3 \vec{k}\,\omega_{\vec{k}}\,a^{~\dagger}_{\vec{k}}\,a_{\vec{k}}~.~\label{e_3.6}
\end{equation}
The interaction Hamiltonian, as the function of proper time, in the interaction picture is found to be
\begin{align}
   &\,\,\tilde{H}_{I}(\tau)\nonumber\\
   &=\mu \int_{-\infty}^{\infty}\,\dfrac{d^3 \vec{k}}{\sqrt{(2\pi)^3\,2\omega_{\vec{k}}}}\bigg[\widetilde{\sigma}_{+}(\tau)\,a_{\vec{k}}\,e^{-i\,k.x(\tau)}+h.c\bigg]\,~\label{e_3.7}
\end{align}
where $k.x(\tau)\equiv\omega_{\vec{k}}\,t(\tau)-\vec{k}\cdot\vec{x}(\tau)$~.

\section{Gorini-Kossakowski-Sudarshan-Lindblad master equation}\label{S:4}
\noindent In this section, we consider the open quantum framework as the system (which is modelled as UDW battery) always interacts with the environment (background quantum field). During the interaction with background quantum field, the battery is moving along the trajectory in Minkowski spacetime with a uniform acceleration and absorbs charging from the external classical pulse. In the laboratory frame, we assume that the trajectory of the moving battery in $(3+1)$-dimensions is given by a worldline function $x(\tau)=(t(\tau),\vec{x}(\tau))$, where $\tau$ is the proper time.

\noindent Following the derivation presented in \cite{mukherjee2024enhancement}, we outline the key steps leading to the Lindblad master equation.\\[3pt]
In the interaction picture $\widetilde{\sigma}_{\pm}(\tau)$ becomes
\begin{equation}\label{e_4.1}
\widetilde{\sigma}_{\pm}(\tau)=e^{iH_{B}{\tau}}\sigma_{\pm}e^{-iH_{B}{\tau}}=s\mp p\,e^{i\alpha\tau}\pm q\,e^{-i\alpha\tau}
\end{equation}
where $s=\sigma_x/2, \,p=(\sigma_z-i\sigma_y)/4,\, q=(\sigma_z+i\sigma_y)/4$.\\[3pt]
The \textit{von Neumann} equation of motion \cite{neu} of the total system is given by 
\begin{equation}\label{e_4.2}
    \frac{\partial\widetilde{\rho}_{tot}(\tau)}{\partial\tau}=-i\Big[\widetilde{H}_{I}(\tau),\widetilde{\rho}_{tot}(\tau)\Big]~.
\end{equation}
The solution of the eq.~\eqref{e_4.2} can be written as
\begin{eqnarray}\label{e_4.3}
    \widetilde{\rho}_{tot}(\tau)=\widetilde{\rho}_{tot}(0)-i\int_{0}^{\tau}du \Big[\widetilde{H}_{I}(u),\widetilde{\rho}_{tot}(u)\Big]~.
\end{eqnarray}
Putting eq.~\eqref{e_4.3} once again in eq.~\eqref{e_4.2} and taking the partial trace over field degrees of freedom, up to $\mathcal{O}(\mu^2)$, we get
\begin{align}\label{e_4.4}
    \frac{\partial\widetilde{\rho}_{s}(\tau)}{\partial\tau}&=-i\,Tr_{\phi}\big[\widetilde{H}_{I}(\tau),\widetilde{\rho}_{tot}(0)\big]\nonumber\\
    &-\int_{0}^{\tau}\d \tau_{1}\,\,Tr_{\phi}\Big[\widetilde{H}_{I}(\tau),\big[\widetilde{H}_{I}(\tau_{1}),\widetilde{\rho}_{tot}(\tau_{1})\big]\Big]~.
\end{align}
Now we consider that at time $t=0$, the total density matrix of the detector-field system is given by
\begin{equation}\label{e_4.5}
    \widetilde{\rho}_{\text{tot}}(0) = \widetilde{\rho}_s(0) \otimes \widetilde\rho_{\phi}(0),
\end{equation}
where $\widetilde{\rho}_s(0)$ is the initial reduced density matrix of the detector, and $\rho_{\phi}(0)=\ketbra{0}{0}$ is the initial reduced density matrix of the field.\\[2pt]
Assuming weak system–environment coupling, the interaction between the system and the environment will not affect the field degrees of freedom, and the field state remains the same at its vacuum state.\\[2pt]
Hence, employing the \textit{Born approximation} \cite{Breuer2007}, we obtain, 
\begin{eqnarray}\label{e_4.6}
    \widetilde{\rho}_{tot}(\tau_{1})\approx \widetilde{\rho}_{s}(\tau_{1})\otimes\rho_{\phi}~.
\end{eqnarray}
At this point, we change the variable $\tau_{1}\rightarrow\tau'$ such that $\tau'=\tau-\tau_{1}$. Therefore, eq.~\eqref{e_4.4} becomes
\begin{align}\label{e_4.7}
    &\frac{\partial\widetilde{\rho}_{s}(\tau)}{\partial\tau}\nonumber\\
    &=-\int_{0}^{\tau}d{\tau'}\,Tr_{F}\Big[\widetilde{H}_{I}(\tau),\big[\widetilde{H}_{I}(\tau-\tau'),\widetilde{\rho}_{s}(\tau-\tau')\otimes\rho_{F}\big]\Big]~.
\end{align}
Using the form of the interaction Hamiltonian given in eq.~\eqref{e_3.7} into eq.~\eqref{e_4.7}, and expanding the double commutator carefully, we can recast eq.~\eqref{e_4.7} as
\begin{align}\label{e_4.8}
    &\frac{\partial\widetilde{\rho}_{s}(\tau)}{\partial\tau}\nonumber\\
&=\mu^2\int_{0}^{\tau}d{\tau'}\bigg[\Big(\widetilde{\sigma}_{-}(\tau-\tau')\widetilde{\rho}_{s}(\tau-\tau')\widetilde{\sigma}_{+}(\tau)\nonumber\\
    &-\widetilde{\sigma}_{+}(\tau)\widetilde{\sigma}_{-}(\tau-\tau')\widetilde{\rho}_{s}(\tau-\tau')\Big)G^{+}(\tau,\tau-\tau')\nonumber\\
    &+\Big(\widetilde{\sigma}_{-}(\tau)\widetilde{\rho}_{s}(\tau-\tau')\widetilde{\sigma}_{+}(\tau-\tau')\nonumber\\
    &-\widetilde{\rho}_{s}(\tau-\tau')\widetilde{\sigma}_{+}(\tau-\tau')\widetilde{\sigma}_{-}(\tau)\Big)\big(G^{+}(\tau,\tau-\tau')\big)^{*}\bigg]~.
\end{align}
The correlation function and its complex conjugate is defined as
\begin{eqnarray}
    G^{+}(\tau,\tau-\tau')&\equiv&\langle 0\vert\phi(x(\tau))\phi(x(\tau-\tau'))\vert 0\rangle\label{e_4.8a}\\
    \big(G^{+}(\tau,\tau-\tau')\big)^{*}&\equiv& \langle 0\vert\phi(x(\tau-\tau'))\phi(x(\tau))\vert 0\rangle~.\label{e_4.8b}
\end{eqnarray}
Now invoking the \textit{Markov approximation} \cite{Breuer2007}, the density matrix \(\widetilde{\rho}_s(\tau - \tau')\) of the system can be replaced by \(\widetilde{\rho}_s(\tau)\). This allows us to extend the upper limit of the integral to infinity without altering its value.\\[2pt]
Therefore, under the \textit{Born-Markov approximation}, eq.~\eqref{e_4.8} can recast as 
\begin{align}\label{e_4.9}
   &\,\,\frac{\partial\widetilde{\rho}_{s}(\tau)}{\partial\tau}\nonumber\\
   &=\mu^2\bigg[\big(q\widetilde{\rho}_{s}(\tau)p-pq\widetilde{\rho}_{s}(\tau)\big)\,\mathcal{I}^{+}_1+\big(p\widetilde{\rho}_{s}(\tau)q-qp\widetilde{\rho}_{s}(\tau)\big)\,\mathcal{I}^{-}_1\nonumber\\
   &+\big(p\widetilde{\rho}_{s}(\tau)q-\widetilde{\rho}_{s}(\tau)qp\big)\,\mathcal{I}^{+}_2+\big(q\widetilde{\rho}_{s}(\tau)p-\widetilde{\rho}_{s}(\tau)pq\big)\,\mathcal{I}^{-}_2\nonumber\\
   &+\big(s\widetilde{\rho}_{s}(\tau)s-s^2\widetilde{\rho}_{s}(\tau)\big)\,\mathcal{I}_3+\big(s\widetilde{\rho}_{s}(\tau)s-\widetilde{\rho}_{s}(\tau)s^2\big)\,\mathcal{I}^{*}_3\bigg]
\end{align}
with
\begin{eqnarray}
\mathcal{I}^{\pm}_1&=&\int_{0}^{\infty}d{\tau'}e^{\pm i \Omega \tau'}\,G^{+}(\tau,\tau-\tau')\label{e_4.10}\\[4pt] 
\mathcal{I}^{\pm}_2&=&\int_{0}^{\infty}d{\tau'}e^{\pm i \Omega \tau'}\,\big(G^{+}(\tau,\tau-\tau')\big)^{*}\label{e_4.11}\\[4pt]
\mathcal{I}_3&=&\int_{0}^{\infty}d{\tau'}\,G^{+}(\tau,\tau-\tau')\label{e_4.12}\\[4pt]
\mathcal{I}^{*}_3&=&\int_{0}^{\infty}d{\tau'}\,\big(G^{+}(\tau,\tau-\tau')\big)^{*}~.\label{e_4.13}
\end{eqnarray}
Substituting the integrals given in eqs.~(\ref{e_4.10}-\ref{e_4.13}) into eq.~\eqref{e_4.9} and after doing some mathematical manipulation, we obtain
\begin{align}\label{e_4.14}
    &\frac{\partial\widetilde{\rho}_{s}(\tau)}{\partial\tau}\nonumber\\
    &=\frac{\mu^2}{2}\Big[\mathcal{G}(\Omega)\big(2q\widetilde{\rho}_{s}(\tau)p-pq\widetilde{\rho}_{s}(\tau)-\widetilde{\rho}_{s}(\tau)pq\big)\nonumber\\
    &\,\,+\mathcal{G}(-\Omega)\big(2p\widetilde{\rho}_{s}(\tau)q-qp\widetilde{\rho}_{s}(\tau)-\widetilde{\rho}_{s}(\tau)qp\big)\nonumber\\
    &\,\,+\mathcal{G}(0)\big(2s\widetilde{\rho}_{s}(\tau)s-s^{2}\widetilde{\rho}_{s}(\tau)-\widetilde{\rho}_{s}(\tau)s^{2}\big)\nonumber\\
    &\,\,+\mathcal{K}(\Omega)\big(\widetilde{\rho}_{s}(\tau)pq-pq\widetilde{\rho}_{s}(\tau)\big)\nonumber\\
    &\,\,+\mathcal{K}(-\Omega)\big(\widetilde{\rho}_{s}(\tau)qp-qp\widetilde{\rho}_{s}(\tau)\big)\nonumber\\
    &\,\,+\mathcal{K}(0)\big(\widetilde{\rho}_{s}(\tau)s^{2}-s^{2}\widetilde{\rho}_{s}(\tau)\big)\Big]~.
\end{align}
where
\begin{eqnarray}
    \mathcal{G}(\pm\Omega)&=&\int_{-\infty}^{+\infty}d{\tau'}e^{\pm i\Omega \tau'}\,G^{+}(\tau,\tau-\tau')~\label{e_4.15}\\[4pt]
    \mathcal{K}(\pm\Omega)&=&\frac{\mathscr{P}}{\pi i}\int_{-\infty}^{+\infty}d{\omega}\frac{\mathcal{G}(\omega)}{\omega\mp\Omega}~\label{e_4.16}
\end{eqnarray}
where $\mathscr{P}$ denotes the \textit{Cauchy principal value}.\\[2pt]
In the Schr\"{o}dinger picture, above eq.~\eqref{e_4.14} turns out to be
\begin{align}\label{e_4.17}
    &\frac{\partial\rho_{s}(\tau)}{\partial\tau}+i\big[H_B,\rho_{s}(\tau)\big]\nonumber\\
    &=\frac{\mu^2}{2}\Big[\mathcal{G}(\Omega)\big(2q\rho_{s}(\tau)p-pq\rho_{s}(\tau)-\rho_{s}(\tau)pq\big)\nonumber\\
    &\,\,+\mathcal{G}(-\Omega)\big(2p\rho_{s}(\tau)q-qp\rho_{s}(\tau)-\rho_{s}(\tau)qp\big)\nonumber\\
    &\,\,+\mathcal{G}(0)\big(2s\rho_{s}(\tau)s-s^{2}\rho_{s}(\tau)-\rho_{s}(\tau)s^{2}\big)\nonumber\\
    &\,\,+\mathcal{K}(\Omega)\big(\rho_{s}(\tau)pq-pq\rho_{s}(\tau)\big)\nonumber\\
    &\,\,+\mathcal{K}(-\Omega)\big(\rho_{s}(\tau)qp-qp\rho_{s}(\tau)\big)\Big]~.
\end{align}
Since $\mathcal{K}(0)\big(\widetilde{\rho}_{s}(\tau)s^{2}-s^{2}\widetilde{\rho}_{s}(\tau)\big)=0$ as $s^{2}\sim \mathds{I}$.\\[2pt]
In terms of Pauli matrices, eq.~\eqref{e_4.17} can be rewritten in the form
\begin{equation}\label{e_4.18}
\frac{\partial \rho_s(\tau)}{\partial \tau}=-i\Big[\mathcal{H}_{eff}, \rho_s(\tau)\Big] + \mathcal{D}\big[\rho_s(\tau)\big]~,
\end{equation}
where the effective Hamiltonian $\mathcal{H}_{eff}$ is given by
\begin{equation}
    \mathcal{H}_{eff}=H_B-\dfrac{i\mu^2}{16}\,\mathcal{K}_{-}(\Omega)\sigma_x~\label{e_4.19}
\end{equation}
and the dissipator $\mathcal{D}\big[\rho_s(\tau)\big]$ takes the form
\begin{align}
&\,\,\mathcal{D}\big[\rho_s(\tau)\big]\nonumber\\
&=\dfrac{\mu^2}{2}\bigg[\dfrac{\mathcal{G}_{+}(\Omega)}{16}\bigg\lbrace \Big(2\sigma_z\rho_s(\tau)\sigma_z-\sigma_z^2\rho_s(\tau)-\rho_s(\tau)\sigma_z^2\Big) \nonumber\\
&+\Big(2\sigma_y\rho_s(\tau)\sigma_y-\sigma_y^2\rho_s(\tau)-\rho_s(\tau)\sigma_y^2\Big) \bigg\rbrace\nonumber\\
&+\dfrac{i\,\mathcal{G}_{-}(\Omega)}{16} \bigg\lbrace \Big(2\sigma_y\rho_s(\tau)\sigma_z-\sigma_z\,\sigma_y\rho_s(\tau)-\rho_s(\tau)\sigma_z\sigma_y\Big) \nonumber\\
&-\Big(2\sigma_z\rho_s(\tau)\sigma_y-\sigma_y\,\sigma_z\rho_s(\tau)-\rho_s(\tau)\sigma_y\sigma_z\Big)   \bigg\rbrace\nonumber\\
&+\dfrac{\mathcal{G}(0)}{4} \bigg\lbrace 2\sigma_x\rho_s(\tau)\sigma_x-\sigma_x^2\rho_s(\tau)-\rho_s(\tau)\sigma_x^2 \bigg\rbrace \bigg]~.\label{e_4.20}
\end{align}
In the above equations, we use the following definitions
\begin{align}
    \mathcal{G}_{\pm}(\Omega)&\equiv\mathcal{G}(\Omega)\pm\mathcal{G}(-\Omega)~\label{e_4.21}\\[4pt]
    \mathcal{K}_{-}(\Omega)&\equiv\mathcal{K}(\Omega)-\mathcal{K}(-\Omega)~.\label{e_4.22}
\end{align}
Eq. \eqref{e_4.18} represents the \textit{GKSL master equation} for the evolution of the reduced density matrix $\rho_{s}(\tau)$ of a linearly coupled Unruh–DeWitt battery interacting with a massless quantum scalar field.

\subsection{Calculations associated with the Wightman function}\label{S:4.1}
\begin{figure}[h!]
\begin{center}
\includegraphics[scale=0.6]{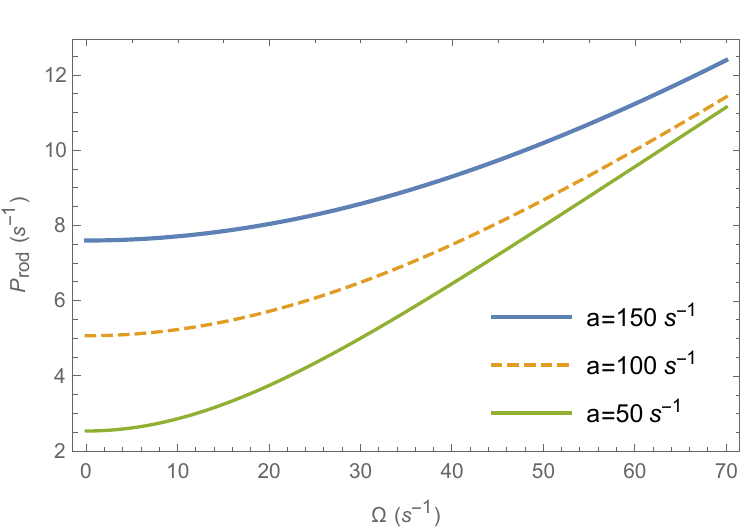}
\caption{Behaviour of the decay rate $P_{rod}$ with frequency $\Omega$ for different numerical values of uniform acceleration $a$.\label{prod}}
\end{center}
\end{figure}

\begin{figure}[h!]
\begin{center}
\includegraphics[scale=0.6]{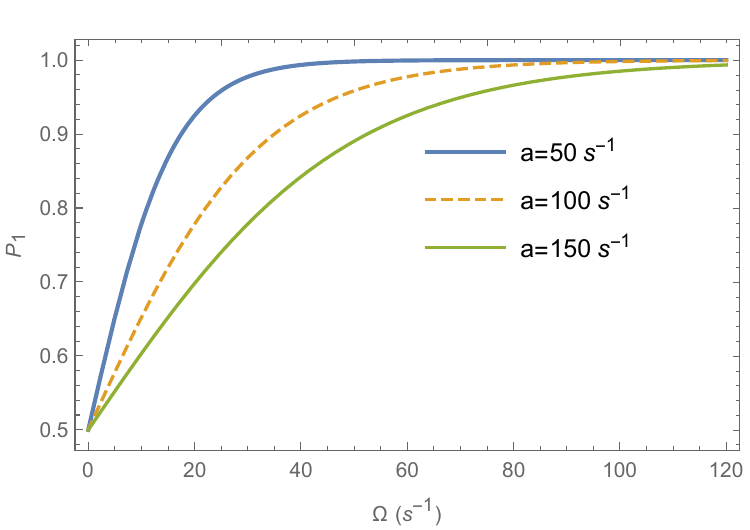}
\caption{Behaviour of $P_{1}$ with frequency $\Omega$ for different numerical values of uniform acceleration $a$. \label{p1}}
\end{center}
\end{figure}

\begin{figure}[h!]
\begin{center}
\includegraphics[scale=0.6]{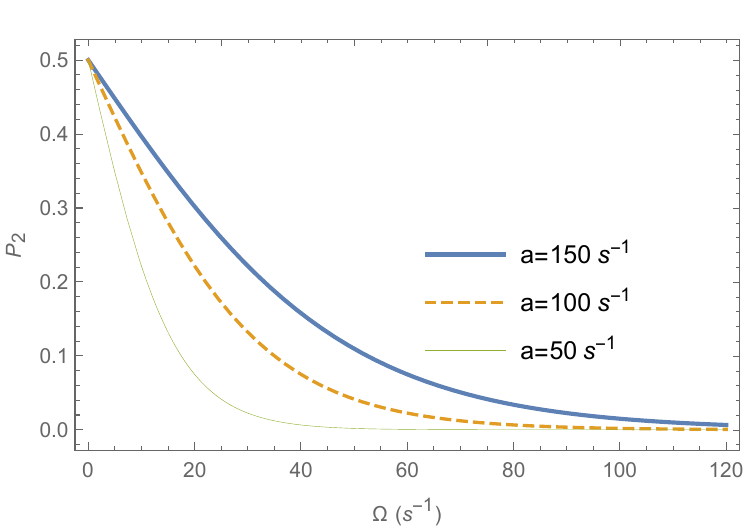}
\caption{Behaviour of $P_{2}$ with frequency $\Omega$ for different numerical values of uniform acceleration $a$.}\label{p2}
\end{center}
\end{figure}

\noindent In this subsection, we are going to compute the Wightman function of the UDW battery by considering that it moves in a worldline, given in the following form \cite{rindler1991introduction}
\begin{equation}
    x(\tau)=\left(\frac{1}{a}\sinh(a\tau),\,\frac{1}{a}\cosh(a\tau),\,0,\,0\right)~.\label{e_4.1.1}
\end{equation}
From the definition given in eq.~\eqref{e_4.8a}, Wightman function is written as 
\begin{equation}
    G^{+}(\tau,\tau-\tau')=\langle 0\vert\phi(x(\tau))\phi(x(\tau-\tau'))\vert 0\rangle~.\label{e_4.1.2}
\end{equation}
For the massless scalar field, above Wightman function can be recast as \cite{birrell1984quantum}
\begin{align}
    &\,\,\,G^{+}(\tau,\tau-\tau')\nonumber\\
    &=-\frac{1}{4\pi^2}\frac{1}{(t(\tau)-t(\tau-\tau')-i\epsilon)^2-\vert \textbf{x}(\tau)-\textbf{x}(\tau-\tau')\vert^2}~.\label{e_4.1.3}
\end{align}
Substituting the trajectory given in eq.~\eqref{e_4.1.1}, the Wightman function along this trajectory turns out to be
\begin{equation}
G^{+}(\tau,\tau-\tau')=-\frac{a^2}{16\pi^2}\left[\sinh^2\left(\frac{a\,\tau'}{2}-i\epsilon\right)\right]^{-1}~.\label{e_4.1.4}
\end{equation}
Putting eq.~\eqref{e_4.1.4} into eq.~\eqref{e_4.15}, we obtain
\begin{equation}
    \mathcal{G}(\Omega)=-\frac{a^2}{16\pi^2}\int_{-\infty}^{+\infty}d{\tau'}e^{i \Omega \tau'}\,\left[\sinh^2\left(\frac{a\,\tau'}{2}-i\epsilon\right)\right]^{-1}~.\label{e_4.1.5}
\end{equation}
By solving the above integral, we get
\begin{equation}
    \mathcal{G}(\Omega)=\frac{\Omega}{2\pi}\frac{e^{2\pi\Omega/a}}{(e^{2\pi\Omega/a}-1)}~.\label{e_4.1.6}
\end{equation}
Through a similar calculation, we also obtain
\begin{equation}
    \mathcal{G}(-\Omega)=\frac{\Omega}{2\pi}\frac{1}{(e^{2\pi\Omega/a}-1)}~.\label{e_4.1.7}
\end{equation}
Taking the limit $\Omega\rightarrow0$, $\mathcal{G}(\Omega)$ becomes
\begin{equation}
    \mathcal{G}(0)=\frac{a}{4\pi^2}~.\label{e_4.1.8}
\end{equation}
Substituting eqs.~\eqref{e_4.1.6} and \eqref{e_4.1.7} in eq.~\eqref{e_4.21}, we obtain
\begin{align}
    \mathcal{G}_{+}(\Omega)&=\frac{\Omega}{2\pi}\,\left[\frac{\exp\Big(\frac{2\pi\Omega}{a}\,\Big)+1}{\exp\Big(\frac{2\pi\Omega}{a}\,\Big)-1}\right]\nonumber\\
    &=\frac{\Omega}{2\pi}\coth\left(\frac{\pi\,\Omega}{a}\right)\nonumber\\&\equiv \textcolor{blue}{P_{rod}}
    ~\label{e_4.1.9}\\[4pt]
    \mathcal{G}_{-}(\Omega)&=\frac{\Omega}{2\pi}~.\label{e_4.1.10}
\end{align}
Here we define $\mathcal{G}_{+}(\Omega)$ to be $P_{rod}$ where the suffix ``rod" signifies ``\textit{rate of decay}" since $\mathcal{G}_{+}(\Omega)$ often appears as the rate at which dissipation occurs in the single time expectation values and in the correlation functions (as we shall see in section \ref{S:5} and \ref{S:6} respectively ) found for the present study.
The point is to note that, although the aforementioned functions provided in eqs. (\ref{e_4.1.6}, \ref{e_4.1.7}, \ref{e_4.1.9}, \ref{e_4.1.10}) are positive only in the domain $\Omega\,>0$, the following ratios associated with those functions always hold positive value irrespective of the value of $\Omega$. Those are found to be
\begin{align}
    \dfrac{\mathcal{G}(\Omega)}{\mathcal{G}_{+}(\Omega)}\,=\,\dfrac{1}{1+e^{-2\pi\Omega/a}}\equiv\,P_{1}(\Omega),\label{e_4.1.10a1}\\
    \dfrac{\mathcal{G}(-\Omega)}{\mathcal{G}_{+}(\Omega)}\,=\,\dfrac{1}{1+e^{2\pi\Omega/a}}\equiv\,P_{2}(\Omega)~.\label{e_4.1.10a}
\end{align}
Analysing the above expressions for any finite value of $\Omega$, it can be concluded that
\begin{align}
0\,<\,\dfrac{\mathcal{G}\big(\pm\Omega\big)}{\mathcal{G}_{+}(\Omega)}\,<\,1 ~. \label{e_4.1.10b}
\end{align}
In the absence of the acceleration ($a=0$) of the battery, the above functions are reduced to be 
\begin{align}
   &\mathcal{G}(\Omega)\Big|_{a=0}=\dfrac{\Omega}{2\pi}~,~\mathcal{G}(-\Omega)\Big|_{a=0}=0~,~\mathcal{G}(0)\Big|_{a=0}=0\label{e_4.1.10c}\\
   &\mathcal{G}_{+}(\Omega)\Big|_{a=0}=\mathcal{G}_{-}(\Omega)\Big|_{a=0}=\dfrac{\Omega}{2\pi}~.\label{e_4.1.10d}
\end{align}
Now, by fixing $\Omega$ as a positive and finite quantity, we intend to analyse Eqs. (\ref{e_4.1.8}, \ref{e_4.1.9}, \ref{e_4.1.10a1}, \ref{e_4.1.10a}) with respect to $a$, the acceleration of the battery. As per Eq. (\ref{e_4.1.8}), $\mathcal{G}(0)$ is linearly proportional to the acceleration.
By observing the simple exponential structures of Eq. (\ref{e_4.1.10a1}, \ref{e_4.1.10a}), it can be readily recognized that both $\mathcal{G}{+}(\Omega)$ and $P_2(\Omega)$ increase with increasing values of $a$. In contrast, $P_1(\Omega)$ decreases as $a$ increases. The behaviour of $P_1(\Omega)$,  $P_2(\Omega)$ ($P_1$ and $P_2$ respectively) and $P_{rod}$, is graphically represented in Figures. (\ref{prod}, \ref{p1}, \ref{p2}), which will be helpful in analysing the nature of the correlation functions in the subsequent section \ref{S:6}.\\

\noindent 

\noindent 
\noindent Putting eq.~\eqref{e_4.1.6}  in eq.~\eqref{e_4.22}, we have 
\begin{align}
    &\,\,\mathcal{K}_{-}(\Omega)\nonumber\\
    &=\frac{1}{2\pi^{2} i}\bigg[\mathscr{P}\int_{-\infty}^{+\infty}d{\omega}\frac{\omega\,e^{2\pi\Omega/a}}{(e^{2\pi\omega/a}-1)(\omega-\Omega)}\nonumber\\
    &\qquad\qquad-\mathscr{P}\int_{-\infty}^{+\infty}d{\omega}\frac{\omega\,e^{2\pi\Omega/a}}{(e^{2\pi\omega/a}-1)(\omega+\Omega)}\bigg]~.\label{e_4.1.11}
\end{align}
After some simplification, we obtain 
\begin{align}
    &\,\,\mathcal{K}_{-}(\Omega)\nonumber\\
    &=\frac{\Omega}{2\pi^{2} i}\bigg[\mathscr{P}\int_{-\infty}^{+\infty}\frac{d\omega}{(e^{2\pi\omega/a}-1)(\omega-\Omega)}\nonumber\\
    &\qquad\qquad+\mathscr{P}\int_{-\infty}^{+\infty}\frac{d{\omega}}{(e^{2\pi\omega/a}-1)(\omega+\Omega)}\bigg]~.\label{e_4.1.12}
\end{align}
Now taking individual integral and solving the integrals using the method of contour integration, we found that both the integrals give the same result, which is
\begin{align}
    &\,\,\mathscr{P}\int_{-\infty}^{+\infty}\frac{d{\omega}}{(e^{2\pi\omega/a}-1)(\omega\pm\Omega)}\nonumber\\
    &=\frac{ia}{2}\sum_{n=1}^{\infty}\left[\frac{1}{(\Omega+ina)}-\frac{1}{(\Omega-ina)} \right]~.\label{e_4.1.13}
\end{align}
Substituting the result of the eq.~\eqref{e_4.1.13} into eq.~\eqref{e_4.1.12} and after simplifying, we obtain
\begin{equation}
    \mathcal{K}_{-}(\Omega)=-\frac{i\Omega}{\pi^2}\sum_{n=1}^{\infty}\frac{n}{n^2+(\frac{\Omega}{a})^2}~.\label{e_4.1.14}
\end{equation}
The form of $\mathcal{K}_{-}(\Omega)$ suggests that it is a divergent quantity. Therefore, we regularise it to extract a finite piece from this. At this point, we consider \textit{exponential regularisation technique} \cite{peskin2018introduction}.\\[2pt]
Hence, we rewrite the eq.~\eqref{e_4.1.14} in the following form
\begin{align}
    &\,\,\mathcal{K}_{-}(\Omega)=-\frac{i\Omega}{\pi^2}\sum_{n=1}^{\infty}\frac{n}{n^2+(\frac{\Omega}{a})^2}\nonumber\\
    &=-\frac{i\Omega}{\pi^2}\lim_{\epsilon\to0}\left[\sum_{n=1}^{\infty}\frac{n\,e^{-\epsilon n}}{n^2+(\frac{\Omega}{a})^2} \right]\nonumber\\
    &=-\frac{i\Omega}{\pi^2}\lim_{\epsilon\to0}\bigg[\frac{1}{2}\sum_{n=1}^{\infty}e^{-\epsilon n}\Big\{\frac{1}{(n-i\frac{\Omega}{a})}+\frac{1}{(n+i\frac{\Omega}{a})}\Big\}\bigg]\nonumber\\
    &=-\frac{i\Omega}{\pi^2}\lim_{\epsilon\to0}\bigg[e^{-\epsilon}\bigg\{\Phi\left(e^{-\epsilon},\,1,\,1-i\frac{\Omega}{a}\right)\nonumber\\
    &\hspace{4cm}+\Phi\left(e^{-\epsilon},\,1,\,1+i\frac{\Omega}{a}\right)\bigg\}\bigg]~\label{e_4.1.15}
\end{align}
where $\Phi\left(e^{-\epsilon},\,1,\,1\mp i\frac{\Omega}{a}\right)$ are the \textit{Hurwitz-Lerch function} \cite{gradshteyn2014table}.\\[2pt]
Now, taking expansion with respect to $\epsilon$, eq.~\eqref{e_4.1.15} becomes 
\begin{align}
    &\,\,\mathcal{K}_{-}(\Omega)\nonumber\\
    &=-\frac{i\Omega}{\pi^2}\lim_{\epsilon\to0}\bigg[2\ln{\left(\frac{1}{\epsilon}\right)}-\psi\left(1+i\frac{\Omega}{a}\right)-\psi\left(1-i\frac{\Omega}{a}\right)\nonumber\\
    &-2\gamma+\mathcal{O}(\epsilon)+\mathcal{O}(\epsilon^2)+\cdots\bigg]~.\label{e_4.1.16}
\end{align}
Here $\psi\left(1\pm i\frac{\Omega}{a}\right)$ are the \textit{digamma function} and $\gamma$ is the \textit{Euler constant} \cite{gradshteyn2014table}.\\[2pt]
Therefore, taking the limit $\epsilon\to0$, the finite regularised part is
\begin{align}
    &\,\,\mathcal{K}_{-}(\Omega)\nonumber\\
    &=-\frac{i\Omega}{\pi^2}\bigg[\psi\left(1+ i\frac{\Omega}{a}\right)+\psi\left(1- i\frac{\Omega}{a}\right)+2\gamma\bigg]\nonumber\\
    &=-\frac{2i\Omega}{\pi^2}\bigg[\Re{\psi\left(1+ i\frac{\Omega}{a}\right)}+\gamma \bigg]~.\label{e_4.1.17}
\end{align}
Substituting eq.~\eqref{e_4.1.17} in eq.~\eqref{e_4.19}, we obtain
\begin{align}
    &\,\,\mathcal{H}_{eff}\nonumber\\
    &=\frac{\Omega}{2}\bigg[1-
    \frac{\mu^2}{4\pi^2}\bigg\{\Re{\psi\left(1+ i\frac{\Omega}{a}\right)}+\gamma \bigg\}\bigg]\sigma_x~.\label{e_4.1.18}
\end{align}
From eq. (\ref{e_4.1.14}), it is quite obvious that $\mathcal{K}_{-}(\Omega)\Big|_{a=0}=0$.


\section{Single time expectation values}\label{S:5}
\noindent In this section, we shall address the dynamical evolution of the operators associated with the system. Since we intend to apply the quantum regression theorem for computing the correlation functions from the single point expectation value of the operators, we shall first fix the form of the ladder operators which are to be defined as per the system Hamiltonian (eq.~\eqref{e_3.5}) of the battery model. \\
\noindent Now, the eigensystem of the system Hamiltonian is  
\begin{equation}
H_B\,\ket{+}\,=\,\Omega\,\ket{+}~;~H_B\,\ket{-}\,=\,-\Omega\,\ket{-}~,
\end{equation}
where the matrix form of the basis $\ket{\pm}$ are
\begin{eqnarray}
\ket{+}=\dfrac{1}{\sqrt{2\,}}
\begin{pmatrix}
1\\
1
\end{pmatrix}~;~
\ket{-}=\dfrac{1}{\sqrt{2\,}}
\begin{pmatrix}
1\\
-1
\end{pmatrix}~.
\end{eqnarray} 
The Pauli operators can be expanded as
\begin{align}
&\sigma_x\,=\,\Big(\ket{+}\bra{+}\Big)-\Big(\ket{-}\bra{-}\Big)~,\label{e_5.3}\\
&\sigma_y\,=\,i\Big[\Big(\ket{+}\bra{-}\Big)-\Big(\ket{-}\bra{+}\Big)\Big]~,\label{e_5.4}\\
&\sigma_z\,=\,\Big(\ket{+}\bra{-}\Big)+\Big(\ket{-}\bra{+}\Big)\label{e_5.5}~.
\end{align}
\subsection{Ladder operators}\label{S:5.1}
\noindent With regard to both the Pauli matrices and the basis $\ket{\pm}$, the ladder operators should be defined in the following way
\begin{eqnarray}
&\sigma^{\,(B)}_{+}\,\equiv\,\dfrac{1}{2}\big[\sigma_z-i\sigma_y\big]\,=\,\ket{+}\bra{-}~,\label{e_5.a.1}\\
& \sigma^{\,(B)}_{-}\,\equiv\,\dfrac{1}{2}\big[\,\sigma_z+i\sigma_y\big]\,=\,\ket{-}\bra{+}~.\label{e_5.a.2}
\end{eqnarray}
The above definition, expectedly, satisfies the following commutation and anti-commutation relations
\begin{eqnarray}
\left[\,\sigma^{\,(B)}_{+}, \sigma^{\,(B)}_{-}\right]\,=\,\sigma_x~;~\left\lbrace\sigma^{\,(B)}_{+}, \sigma^{\,(B)}_{-}\right\rbrace\,=\,\mathcal{1}~;
\end{eqnarray}
and perform the raising and lowering operations as
\begin{equation}
\sigma^{\,(B)}_{+}\,\ket{-}\,=\,\ket{+}~,~\sigma^{\,(B)}_{-}\,\ket{+}\,=\,\ket{-}~.
\end{equation}
It is also very straightforward to verify that
\begin{equation}
\big({\sigma^{\,(B)}_{\pm}}\big)^2\,=\,0~,~\big({\sigma^{\,(B)}_{\pm}\sigma^{\,(B)}_{\mp}}\big)^2\,=\,\sigma^{\,(B)}_{\pm}\sigma^{\,(B)}_{\mp}~.
\end{equation}
Next, we aim to establish the dynamical differential equation governed by the time dependent expectation values of these 
operators. Those can be easily derived from the matrix element of the GKSL master equation (eq.~\eqref{e_4.18}). Apart from this, one can also deduce the same from the dynamic evolution of the expectation value of the Pauli operators. In both ways, we would proceed and show the consistency between the two methods. 
\subsection{Matrix elements of GKSL master equation}\label{S:5.2}
\noindent The master equation (eq.\ref{e_4.18}) can be simplified in the following form.
\begin{align}
&\dot{\rho}_{s}(\tau)\,=\,\Big(i\,\Omega+\dfrac{\mu^2}{16}\mathcal{K}_{-}(\Omega)\Big)\,\big[\,\rho_{s}(\tau),\,\sigma_x \big]\nonumber\\
&+\dfrac{\mu^2}{16}\Bigg[\mathcal{G}_{+}(\Omega)\Big\lbrace\sigma_z\,\rho_{s}(\tau)\,\sigma_z+\sigma_y\,\rho_{s}(\tau)\,\sigma_y-2\,\rho_{s}(\tau)\Big\rbrace\nonumber\\
&+i\,\mathcal{G}_{-}(\Omega)\Big\lbrace \sigma_y\,\rho_{s}(\tau)\,\sigma_z-\sigma_z\,\rho_{s}(\tau)\,\sigma_y+i\,\big\lbrace\,\rho_{s}(\tau),\,\sigma_x \big\rbrace\Big\rbrace\Bigg] \nonumber\\
&+\dfrac{\mu^2\,\mathcal{G}(0)}{4}\Big(\sigma_x\,\rho(\tau)\,\sigma_x-\rho_{s}(\tau)\Big)~.\label{e_5_B.1}
\end{align}
After considering $\mathcal{K}_{-}(\Omega)\,=\,i\,K_{-}(\Omega)$, the following matrix elements of the master equation easily emerge after substituting eqs.~(\ref{e_5.3}, \ref{e_5.4}, \ref{e_5.5}) in the above equation. Hence, we have
\begin{align}
{\dot{\rho}}_{++}&=-\dfrac{\mu^2}{8}\Big[\,\mathcal{G}_{+}(\Omega)\,\big(\,\rho_{++}-\rho_{--}\,\big)+\mathcal{G}_{-}(\Omega)\Big]\,,\label{e_5_B.2}\\
{\dot{\rho}}_{+-}&=-\Bigg[\dfrac{\mu^2}{2}\Big(\mathcal{G}(0)+\dfrac{\mathcal{G}_{+}(\Omega)}{4}\Big)+i\Big(2\,\Omega+\dfrac{\mu^2}{8}K_{-}(\Omega)\Big)  \Bigg]\,\rho_{+-}\,,\label{e_5_B.3}\\
{\dot{\rho}}_{-+}&=-\Big[\dfrac{\mu^2}{2}\Big(\mathcal{G}(0)+\dfrac{\mathcal{G}_{+}(\Omega)}{4}\Big)-i\Big(2\,\Omega+\dfrac{\mu^2}{8}K_{-}(\Omega)\Big)\Big]\,\rho_{-+}\,,\label{e_5_B.4}\\
{\dot{\rho}}_{--}&=\dfrac{\mu^2}{8}\Big[\,\mathcal{G}_{+}(\Omega)\,\big(\,\rho_{++}-\rho_{--}\,\big)+\mathcal{G}_{-}(\Omega)\Big]~.\label{e_5_B.5}
\end{align}
Observing the above equations, one can immediately note that 
${\dot{\rho}}_{++}+{\dot{\rho}}_{--}\,=\,0$. This is a reflection of the fact that the trace of the density matrix is unity. Moreover, it is also realized that the generic form of the cross diagonal elements ${\dot{\rho}}_{+-}$ and ${\dot{\rho}}_{-+}$ should be 
\begin{align}
\rho_{+-}(\tau)
&=\rho_{+-}(0)\,e^{-D_1\tau}~,\label{e_5_B.6}\\
\rho_{-+}(\tau)&=\rho_{-+}(0)\,e^{-D_2\tau}\label{e_5_B.7},
\end{align}
where $D_1$, and $D_2$ are given by
\begin{align}
    D_1&=\dfrac{\mu^2}{2}\Big[\mathcal{G}(0)+\dfrac{\mathcal{G}_{+}(\Omega)}{4}\Big]+i\Big[2\Omega+\dfrac{\mu^2}{8}K_{-}({\Omega})\Big]~,\label{e_5_B.7.1}\\
    D_2&=\dfrac{\mu^2}{2}\Big[\mathcal{G}(0)+\dfrac{\mathcal{G}_{+}(\Omega)}{4}\Big]-i\Big[2\Omega+\dfrac{\mu^2}{8}K_{-}({\Omega})\Big]~.\label{e_5_B.7.2}
\end{align}
However, both matrix elements (coherence terms) are finally revealed to be zero because of the consideration that the initial state of the battery system was at $\ket{+}$. That implies $\rho_{s}(0)\,=\,\ket{+}\bra{+}$. Therefore, 
\begin{align}
\rho_{+-}(0)&=\braket{+}{+}\braket{+}{-}=0~\nonumber\\
\Rightarrow\,\rho_{+-}(\tau)&=0~,\label{e_5_B.8}\\
\rho_{-+}(0)&=\braket{-}{+}\braket{+}{+}=0~\nonumber\\
\Rightarrow\,\rho_{-+}(\tau)&=0~.\label{e_5_B.9}
\end{align}
In principle, one can also find out the expression of the diagonal matrix elements (population terms) by solving eq. (\ref{e_5_B.2}) and eq. (\ref{e_5_B.5}). \\
\noindent Next, according to the relation $\big<\dot{\mathcal{O}}\big>\,=\,Tr\,\big[\mathcal{O}\,\dot{\rho}\,\big]$, the dynamical evolution equation of the expectation value of the operators that are relevant to study the system can be immediately found in terms of the above matrix elements. Since, the consideration $\rho_{s}(0)\,=\,\ket{+}\bra{+}$ reduces the integration constant to be zero, we can directly write 
\begin{align}
\big<\sigma^{(B)}_{-}\big>\,=\,\rho_{+-}~&,~\big<\sigma^{(B)}_{+}\big>\,=\,\rho_{-+}~,\label{e_5_B.10}\\
\big<\sigma_x\big>\,=\,\rho_{++}-\rho_{--}~&,~\big<\sigma^{(B)}_{+}\sigma^{(B)}_{-}\big>\,=\,\rho_{++}~;\label{e_5_B.11}
\end{align}
where all the quantities in above equations are function of the proper time $\tau$. 
\subsection{Expectation value : Pauli operators}\label{S:5.3}
\noindent Since, we intend to explore an alternative way to derive our sought average value of the ladder operators and number operator, we shall first calculate the expectation value of the Pauli operators. By using eq. (\ref{e_5_B.1}), the equation of motion of the Pauli operators are found to be  
\begin{align}
\dot{\expval{\sigma_x}}&=-\dfrac{\mu^2}{4}\Big[\,\mathcal{G}_{+}(\Omega)\,\expval{{\sigma_x}}+\mathcal{G}_{-}(\Omega) \Big]~,\label{e_5_c.1}\\
\dot{\expval{\sigma_y}}&=-2\Big(\Omega-\dfrac{i\mu^2}{16}\mathcal{K}_{-}({\Omega})\Big)\expval{\sigma_z}\nonumber\\
&-\dfrac{\mu^2}{2}\Big(\mathcal{G}(0)+\dfrac{\mathcal{G}_{+}(\Omega)}{4}\Big)\expval{\sigma_y}~,\label{e_5_c.2} \\
\dot{\expval{\sigma_z}}&=2\Big(\Omega-\dfrac{i\mu^2}{16}\mathcal{K}_{-}({\Omega})\Big)\expval{\sigma_y}\nonumber\\
&-\dfrac{\mu^2}{2}\Big(\mathcal{G}(0)+\dfrac{\mathcal{G}_{+}(\Omega)}{4}\Big)\expval{\sigma_z}~\label{e_5_c.3}.
\end{align} 
The initial assumption $\rho_{s}(0)\,=\,\ket{+}\bra{+}$ also fix the following average value of the operators at $\tau\,=\,0$.  
\begin{equation}
{\big<\sigma_x\big>}_{\tau=0}\,=\,1\,,\,{\big<\sigma_y\big>}_{\tau=0}\,=\,0\,,\,{\big<\sigma_z\big>}_{\tau=0}\,=\,0~\label{e_5_c.4}.
\end{equation}
Next, we combine eq. (\ref{e_5_c.2}) and eq. (\ref{e_5_c.3}) in the following matrix representation
\begin{equation}
\begin{pmatrix}
\dot{\expval{\sigma_y}}\\
\dot{\expval{\sigma_z}}
\end{pmatrix}\,=\,
{\mathcal{A}}_{(2\times 2)}\,
\begin{pmatrix}
\expval{\sigma_y}\\
\expval{\sigma_z}
\end{pmatrix}~;\label{e_5_c.5}
\end{equation}
where
\begin{equation}
\mathcal{A}\,=\,
\begin{pmatrix}
-\dfrac{\mu^2}{2}\Big[\mathcal{G}(0)+\dfrac{\mathcal{G}_{+}(\Omega)}{4}\Big]&-2\Big[\Omega-\dfrac{i\mu^2}{16}\mathcal{K}_{-}({\Omega})\Big]\\\\
2\Big[\Omega-\dfrac{i\mu^2}{16}\mathcal{K}_{-}({\Omega})\Big]&-\dfrac{\mu^2}{2}\Big[\mathcal{G}(0)+\dfrac{\mathcal{G}_{+}(\Omega)}{4}\Big]
\end{pmatrix}~\label{e_5_c.6}.
\end{equation}
The general solution of the above equation is as follows
\begin{equation}
\begin{pmatrix}
\expval{\sigma_y}\\
\expval{\sigma_z}
\end{pmatrix}=C_1\,e^{D_3 \tau}\begin{pmatrix}
1\\
i
\end{pmatrix}\,
+C_2\,e^{D_4 \tau}\begin{pmatrix}
1\\
-i
\end{pmatrix}~;\label{e_5_c.7}
\end{equation}
where $C_1$, $C_2$ are constants and $D_3$, $D_4$ are given by
\begin{align}
    D_3&=\dfrac{\mu^2}{2}\Big[\mathcal{G}(0)+\dfrac{\mathcal{G}_{+}(\Omega)}{4}\Big]+i\Big[2\Omega-\dfrac{i\mu^2}{8}\mathcal{K}_{-}({\Omega})\Big]~,\label{e_5_c.7.1}\\
    D_4&=\dfrac{\mu^2}{2}\Big[\mathcal{G}(0)+\dfrac{\mathcal{G}_{+}(\Omega)}{4}\Big]-i\Big[2\Omega-\dfrac{i\mu^2}{8}\mathcal{K}_{-}({\Omega})\Big]~.\label{e_5_c.7.2}
\end{align}
The value of      
$\big<\sigma_y\big>$ and $\big<\sigma_z\big>$, at $\tau\,=\,0$, provide
\begin{eqnarray}
C_1+C_2\,=\,0\,,\label{e_5_c.8}\\
i\big(C_1-C_2\big)\,=\,0~.\label{e_5_c.9}
\end{eqnarray}
Hence, $C_1\,=\,C_2\,=\,0$ and result in zero average value of those operators at any arbitrary proper time $\tau$.  However, by integrating the left hand side of eq. (\ref{e_5_c.1}) and implementing $\big<\sigma_x\big>_{\tau=0}\,=\,1$, a non-zero dynamical expression of $\big<\sigma_x\big>$ is obtained. Therefore, we finally have
\begin{align}
\expval{\sigma_x}&=\dfrac{1}{{\mathcal{G}}_{+}(\Omega)}\,\Big[-{\mathcal{G}}_{-}(\Omega)+2\,\mathcal{G}(\Omega)\,e^{-\frac{\mu^2}{4}{\mathcal{G}}_{+}(\Omega)\,\tau}\,\Big],\label{e_5_c.10}\\
\expval{\sigma_y}&=\expval{\sigma_z}=0~.\label{e_5_c.11}
\end{align}
With the aforementioned results in this section [\ref{S:5.2}] and in the previous section [\ref{S:5.3}], we are now ready to construct the framework which can allow us to implement the quantum regression theorem.
\subsection{Expectation value : Ladder operators and number operator}\label{S:5.4}
\noindent In this subsection, we proceed to establish the set of differential equations obeyed by the expectation values of the ladder operators and the number operator. By solving those equations, we will also show that the resultant expectation values are consistent with the density matrix elements found \big(eqs. [\ref{e_5_B.10}, \ref{e_5_B.11}]\big) and the expectation value of the Pauli operators in the earlier sections.\\ 
\noindent Taking into account eqs. (\ref{e_5.a.1}, \ref{e_5.a.2}) at first, both the set of equations given in eqs. (\ref{e_5_B.10}, \ref{e_5_B.11}) and in eqs. (\ref{e_5_c.1}, \ref{e_5_c.2}, \ref{e_5_c.3}), finally yield 
\begin{align}
   \begin{pmatrix}
\expval{\dot{\sigma}_{-}^{\,(B)}}\\\\
\expval{\dot{\sigma}_{+}^{\,(B)}}\\\\
\expval{\dot{\sigma_{+}^{(B)}\sigma_{-}^{(B)}}}
\end{pmatrix}\,=\, &\begin{pmatrix}
M_{11}&0&0\\\\\\
0&M_{22}&0\\\\\\
0&0&M_{33}
\end{pmatrix}\,\,\begin{pmatrix}
\expval{\sigma_{-}^{\,(B)}}\\\\
\expval{\sigma_{+}^{\,(B)}}\\\\
\expval{\sigma_{+}^{(B)}\sigma_{-}^{(B)}}
\end{pmatrix}\nonumber\\
+&\begin{pmatrix}
0\\
0\\
\dfrac{\mu^2}{4}\,\mathcal{G}(-\Omega)
\end{pmatrix}~;\label{e_5.d.1}
\end{align}
where
\begin{align}
&\,M_{11}\,=\,-\dfrac{\mu^2}{2}\,\Big(\mathcal{G}(0)+\dfrac{\mathcal{G}_{+}(\Omega)}{4} \Big)-\,i\,\Big(2\,\Omega+\dfrac{\mu^2}{8}K_{-}(\Omega) \Big)\,,\label{e_5.d.2}\\
&\,M_{22}\,=\,-\dfrac{\mu^2}{2}\,\Big(\mathcal{G}(0)+\dfrac{\mathcal{G}_{+}(\Omega)}{4} \Big)+\,i\,\Big(2\,\Omega+\dfrac{\mu^2}{8}K_{-}(\Omega) \Big)\,,\label{e_5.d.3}\\
&\,M_{33}\,=\,-\dfrac{\mu^2}{4}\,\mathcal{G}_{+}(\Omega)~.\label{e_5.d.4}
\end{align}
Moreover, for the sake of showing consistency, by using $\sigma_x\,=\,1-2\sigma_{-}^{(B)}\sigma_{+}^{(B)}$, we also deduce
\begin{equation}
\dot{\expval{\sigma_{-}^{(B)}\sigma_{+}^{(B)}}}\,=\,-\dfrac{\mu^2}{4}\,\mathcal{G}_{+}(\Omega)\,\expval{\sigma_{-}^{(B)}\sigma_{+}^{(B)}}\,+\,\dfrac{\mu^2}{4}\mathcal{G}(\Omega)~.\label{e_5.d.5}
\end{equation}
Typically, we find that
\begin{align}
\dot{\expval{\sigma_{+}^{(B)}\sigma_{-}^{(B)}}}+\dot{\expval{\sigma_{-}^{(B)}\sigma_{+}^{(B)}}}\,=\,0~.\label{e_5.d.6}
\end{align}
For simplicity of illustration, we present eq. (\ref{e_5.d.1}) in the  following vector form,
\begin{equation}
\dot{\expval{\boldsymbol{\mathcal{S}}}}\,=\,\boldsymbol{M}\,\expval{\boldsymbol{\mathcal{S}}}\,+\,\boldsymbol{\mathcal{C}}\,;\label{e_5.d.7}
\end{equation}
where 
\begin{align}
\boldsymbol{\mathcal{S}}\,\equiv\,
\begin{pmatrix}
\sigma_{-}^{\,(B)}\\\\
\sigma_{+}^{\,(B)}\\\\
\sigma_{+}^{(B)}\sigma_{-}^{(B)}
\end{pmatrix}\,,\,
\boldsymbol{\mathcal{C}}\,\equiv\,
\begin{pmatrix}
0\\\\
0\\\\
\dfrac{\mu^2}{4}\mathcal{G}(-\Omega)
\end{pmatrix}\,;\label{e_5.d.8}
\end{align}
and $\boldsymbol{M}\,\equiv$ diag $\Big(M_{11},~M_{22},~\,M_{33}\,\Big)$.\\
\noindent Now, we proceed to find  the solution of eq. (\ref{e_5.d.1}) or eq. (\ref{e_5.d.7}). Once again, by using $\rho_{s}(0)\,=\,\ket{+}\bra{+}$, we have the following average value of the operators at $\tau\,=\,0$,  
\begin{align}
&\big<\sigma_{-}^{(B)}\big>_{\tau=0}=0~,\quad\big<\sigma_{+}^{(B)}\big>_{\tau=0}=0~,\label{e_5.d.9}\\
&\big<\sigma_{+}^{(B)}\sigma_{-}^{(B)}\big>_{\tau=0}\,=\,1~,\quad\big<\sigma_{-}^{(B)}\sigma_{+}^{(B)}\big>_{\tau=0}\,=\,0~;\label{e_5.d.10}
\end{align}
which are, expectedly, in accordance with eq. (\ref{e_5_c.4}). \\
\noindent Finally, the formal solution set for the time dependent expectation values of ladder and number operators is given by,
\begin{align}
&\big<\sigma_{-}^{(B)}\big>\,=\,0~,\quad\big<\sigma_{+}^{(B)}\big>\,=\,0~;\\
&\big<\sigma_{+}^{(B)}\sigma_{-}^{(B)}\big>\,=\,\dfrac{1}{\mathcal{G}_{+}(\Omega)}\Big[\mathcal{G}(-\Omega)+\mathcal{G}(\Omega)\,e^{-\frac{\mu^2}{4}\mathcal{G}_{+}(\Omega)\tau} \Big]\,.\label{e_5.d.11}
\end{align}
We also derive, from eq. (\ref{e_5.d.5}), that
\begin{align}
\big<\sigma_{-}^{(B)}\sigma_{+}^{(B)}\big>\,=\,\frac{\mathcal{G}(\Omega)}{\mathcal{G}_{+}(\Omega)}\Big[\,1-\,e^{-\frac{\mu^2}{4}\mathcal{G}_{+}(\Omega)\tau} \Big]~,\label{e_5.d.12}
\end{align}
to verify that they are $\big<\sigma_{+}^{(B)}\sigma_{-}^{(B)}\big>+\big<\sigma_{-}^{(B)}\sigma_{+}^{(B)}\big>\,=\,1$ always. At this point, it is relevant to explain the physical significance of the results shown in eqs. (\ref{e_5.d.9}–\ref{e_5.d.12}).\\[1pt]
\noindent Since, $\sigma_{+}^{(B)}\,=\,\ket{+}\bra{-}\,=\,\rho_{-+}$ and $\sigma_{-}^{(B)}\,=\,\ket{-}\bra{+}\,=\,\rho_{+-}$, the off diagonal elements of the density matrix are essentially coherence (transition) operators. They quantify the quantum superposition and phase coherence between the two states. The condition $\big<\sigma_{\pm}^{(B)}(\tau)\big>\,=\,0$ implies the absence of single-time coherence. This means that there is no coherent superposition or phase coherence between the states $\ket{+}$ and $\ket{-}$. Consequently, the system does not emit coherent radiation, which would require a constant phase relationship, and is typically achieved through stimulated emission. This situation naturally arises when the system starts from a pure state, whether it is the ground state or the excited state. However, the system can still be excited or dynamically active and may emit spontaneous radiation, in which the emitted photons are not phase coherent.\\
\noindent The quantity $\big<\sigma_{+}^{(B)}\sigma_{-}^{(B)}\big>(\tau)$ represents the projector onto the excited state and quantifies the probability of finding the system in the excited state at time $\tau$. In other words, it describes how much excitation is stored in the system at that instant. Initially, the probability of being in the excited state starts from its value of unity and decays with time for finite value of $\Omega\,>\,0$. In the long-time limit, it approaches a positive, finite value of $\mathcal{G}(-\Omega)\big/\mathcal{G}_{+}(\Omega)$. In contrast, the probability of being in the ground state is zero at the initial time; then it increases exponentially for finite value of $\Omega\,>\,0$ and eventually saturates to a positive, finite value of $\mathcal{G}(\Omega)\big/\mathcal{G}_{+}(\Omega)$ at long times. Since the system energy is proportional to $\big<\sigma_x\big>$, the decaying nature of $\big<\sigma_{+}^{(B)}\sigma_{-}^{(B)}\big>(\tau)$ indicates energy loss to the environment through spontaneous emission.\\
\noindent Here we intend to mention a subtle and notable observation regarding this decay. As evident from eqs. (\ref{e_4.1.9}, \ref{e_4.1.10d}), the rate of decay $\mathcal{G}_{+}(\Omega)$ depends on $a$, the acceleration of the battery. For a fixed value of $\Omega$, if we regulate the value of uniform acceleration, eq. (\ref{e_4.1.9}) tells us that the rate of decay increases with the increasing value of $a$ and in the limit $a=0$, $\mathcal{G}_{+}(\Omega)$ takes the lowest value (provided in eq. (\ref{e_4.1.10d})) for a fixed value of $\Omega$. This is also presented in Figure. (\ref{prod}). Hence, dissipation is proportional to the value of the uniform acceleration.\\[3pt]
\noindent Now, we have all the time dependent expectation values required to proceed our study further. Our next job is to ensure the proper framework for implementation of quantum regression theorem for calculating the correlation functions by using eq. (\ref{e_5.d.7}). To do so, eq. (\ref{e_5.d.7}) is required to be represented in homogeneous form. Hence, we can rewrite eq. (\ref{e_5.d.7}) as
\begin{equation}
\dfrac{d}{d\tau}\expval{\boldsymbol{\mathcal{S}}\,+\,\boldsymbol{M}^{-1}\,\boldsymbol{\mathcal{C}}}\,=\,\boldsymbol{M}\,\expval{\boldsymbol{\mathcal{S}}\,+\,\boldsymbol{M}^{-1}\,\boldsymbol{\mathcal{C}}}\label{e_5.d.13}
\end{equation}
and can also define 
\begin{align}
\delta\,\boldsymbol{\mathcal{S}}\,\equiv\,\boldsymbol{\mathcal{S}}\,+\,\boldsymbol{M}^{-1}\,\boldsymbol{\mathcal{C}}~.\label{e_5.d.14}
\end{align}
Since $\boldsymbol{M}^{-1}\,\boldsymbol{\mathcal{C}}$ is a time-independent quantity, therefore, differentiating eq. \eqref{e_5.d.14} gives $\dot{\big<\delta\,\boldsymbol{\mathcal{S}}\big>}\,=\,\dot{\big<\boldsymbol{\mathcal{S}}\big>}$. As also explained in \cite{carmichael1999statistical}, it is actually the expectation value of the operator $\boldsymbol{\mathcal{S}}$ when it reaches the steady state and 
$\dot{\big<\boldsymbol{\mathcal{S}}\big>}_{ss}=\,0$. It yields
\begin{align}
\big<\boldsymbol{\mathcal{S}}\big>_{ss}\,=\,-\boldsymbol{M}^{-1}\,\boldsymbol{\mathcal{C}}~;\label{e_5.d.15}
\end{align}
and, from eq. (\ref{e_5.d.14}), we have
\begin{align}
\delta\,\boldsymbol{\mathcal{S}}\,\equiv\,\boldsymbol{\mathcal{S}}\,-\,\big<\boldsymbol{\mathcal{S}}\big>_{ss}\,=\,\boldsymbol{\mathcal{S}}\,+\,\boldsymbol{M}^{-1}\,\boldsymbol{\mathcal{C}}~.\label{e_5.d.16}
\end{align}
It is noteworthy to mention that the above equations describe the fluctuations about steady state. It basically describes how far the system is from it's equilibrium or steady state and these fluctuations are intrinsic characteristics of quantum dynamics.\\
The individual components of the above vector equation are found to be 
\begin{align}
&\delta\,\sigma_{\pm}^{(B)}\,\equiv\,\sigma_{\pm}^{(B)}~,\label{e_5.d.17}\\
&\delta\sum\,\equiv\,\sum\,-\,\dfrac{\mathcal{G}(-\Omega)}{\mathcal{G}_{+}(\Omega)}~;\label{e_5.d.18}.
\end{align}
where $\sum\equiv\,\sigma_{+}^{(B)}\sigma_{-}^{(B)}$.
Therefore, the most simplified homogeneous form of eq. (\ref{e_5.d.7}) is 
\begin{equation}
\dot{\expval{\delta\boldsymbol{\mathcal{S}}}}\,=\,\boldsymbol{M}\,\expval{\delta\boldsymbol{\mathcal{S}}}\,;\label{e_5.d.19}
\end{equation}
which, in matrix representation, is given by 
\begin{align} 
\begin{pmatrix}
\big<\dot{\delta\,\sigma}_{-}^{\,(B)}\big>\\\\
\big<\dot{\delta\,\sigma}_{+}^{\,(B)}\big>\\\\
\big<\dot{\delta\sum}\big>\\
\end{pmatrix}\,=\,&\begin{pmatrix}
M_{11}&0&0\\\\
0&M_{22}&0\\\\
0&0&M_{33}
\end{pmatrix}\,\begin{pmatrix}
\big<\delta\,\sigma_{-}^{\,(B)}\big>\\\\
\big<\delta\,\sigma_{+}^{\,(B)}\big>\\\\
\big<\delta\sum\big>\\
\end{pmatrix}~;\label{e_5.d.20}
\end{align}
where the matrix elements of $\boldsymbol{M}$ are already provided in eqs. (\ref{e_5.d.2}-\ref{e_5.d.4}).\\
\noindent In an effort to verify consistency, defining $\Delta\equiv\,\sigma_{-}^{(B)}\sigma_{+}^{(B)}$ and applying $\dot{\delta\Delta}\,=\,0$ in eq. (\ref{e_5.d.5}), we also find out
\begin{equation}
\delta\Delta\,\equiv\,\Delta\,-\,\dfrac{\mathcal{G}(\Omega)}{\mathcal{G}_{+}(\Omega)}~;\label{e_5.d.21}
\end{equation}
which obeys
\begin{equation}
\big<\dot{\delta\Delta}\big>\,=\,-\dfrac{\mu^2}{4}\,\mathcal{G}_{+}(\Omega)\,\big<\delta\Delta\big>~.\label{e_5.d.22}
\end{equation}
It is very easy to check that both
\begin{equation}
\expval{\delta\Delta}+\expval{\delta\sum}\,\,=\,\expval{\dot{\delta\Delta}}+\expval{\dot{\delta\sum}}\,=\,0~;\label{e_5.d.23}
\end{equation}
which is actually consistent with $\big<\sigma_{+}^{(B)}\sigma_{-}^{(B)}\big>+\big<\sigma_{-}^{(B)}\sigma_{+}^{(B)}\big>\,=\,1$. \\
\noindent In next section, we shall compute two-time correlation functions from eq. (\ref{e_5.d.19} or \ref{e_5.d.20}) by directly imposing the quantum regression theorem on it.

\section{Quantum Regression Theorem\,:\,Two-time correlation functions}\label{S:6}
\noindent This section is entirely devoted to the application of the quantum regression theorem. Using this theorem, we aim to derive several examples of non-zero correlation functions, which will later lead to results describing significant physical effects associated with the relativistic quantum battery model. Although we attempt to analyse all possible non-zero first-order correlation functions, we restrict our discussion of second-order correlations to a specific example which corresponds to a physical phenomenon. 
\subsection{First-order correlation functions}\label{S:6.1}
\noindent As also shown in eq. (\ref{e.2.3}), the first-order correlation function, for our context, is essentially a two-time correlation function which measures the dependence between two successive operations performed at two different points in time, where the later operation gets influenced by the earlier one.\\
\noindent We begin by mentioning the vector form of the equation of motion for the first-order correlation function based on the same for the time dependent expectation value found in the earlier section. If
\begin{align}
\big<\dot{\delta\boldsymbol{\mathcal{S}}}\big>\,=\,\boldsymbol{M}\,\big<\delta\boldsymbol{\mathcal{S}}\big>~,\label{e.6.a.1}
\end{align}
then according to the quantum regression theorem, the equation of motion for the probability that the operation $\delta\,\boldsymbol{\mathcal{S}}(\tau^{\prime} +\tau)$ performed at $(\tau^{\prime} +\tau)$, is affected by an earlier operation $\boldsymbol{\mathcal{O}}(\tau^{\prime})$ performed at $\tau^{\prime}$, is given by
\big($\tau\,\geq\,0$\big) 
\begin{align}
\dfrac{d}{d\tau}\bigg<\mathcal{O}(\tau^{\prime})\,\delta\,\boldsymbol{\mathcal{S}}(\tau+\tau^{\prime})\bigg>=\boldsymbol{M}\,\bigg<\mathcal{O}(\tau^{\prime})\,\delta\,\boldsymbol{\mathcal{S}}(\tau+\tau^{\prime})\bigg>~;\label{e.6.a.2}
\end{align}
where $\mathcal{O}$ can be any operator acting in the Hilbert space of the system.  
In terms of matrix element, the above equation takes the following form \big(for $\mu\in[1,3]$\big) 
\begin{align}
\dfrac{d}{d\tau}\bigg<\mathcal{O}(\tau^{\prime})\,\delta\,\mathcal{S}_{\mu}(\tau+\tau^{\prime})\bigg>=\sum_{\lambda=1}^{3}M_{\mu\lambda}\,\bigg<\mathcal{O}(\tau^{\prime})\,\delta\,\mathcal{S}_{\lambda}(\tau+\tau^{\prime})\bigg>\,.\label{e.6.a.3}
\end{align}
With regard to the specific forms of the operator $\mathcal{O}$, eq. (\ref{e.6.a.2}) can be rewritten as
\begin{align}
&\dot{\big<\delta\,\sigma_{-}^{(B)}(\tau^{\prime})\,\delta\,\boldsymbol{\mathcal{S}}(\tau+\tau^{\prime})\big>}=\boldsymbol{M}\,\big<\delta\,\sigma_{-}^{(B)}(\tau^{\prime})\,\delta\,\boldsymbol{\mathcal{S}}(\tau+\tau^{\prime})\big>~;\label{e.6.a.4}\\
&\dot{\Big<\delta\,\sigma_{+}^{(B)}(\tau^{\prime})\,\delta\,\boldsymbol{\mathcal{S}}(\tau+\tau^{\prime})\big>}=\boldsymbol{M}\,\big<\delta\,\sigma_{+}^{(B)}(\tau^{\prime})\,\delta\,\boldsymbol{\mathcal{S}}(\tau+\tau^{\prime})\big>~;\label{e.6.a.5}\\
&\dot{\Big<\delta\,\sum(\tau^{\prime})\,\delta\,\boldsymbol{\mathcal{S}}(\tau+\tau^{\prime})\Big>}=\boldsymbol{M}\,\Big<\delta\,\sum(\tau^{\prime})\,\delta\,\boldsymbol{\mathcal{S}}(\tau+\tau^{\prime})\Big>~;\label{e.6.a.6}
\end{align}
which can also be addressed as the fluctuation correlation functions.\\
\noindent Thus, there are nine total correlation functions of first order that can be obtained by solving the above equations. However, most of them are revealed to be zero based on our chosen initial condition that is $\rho_s(0)\,=\,\ket{+}\bra{+}$ and three non zero solutions would be provided from each of the above equations.\\
\noindent We first address eq. (\ref{e.6.a.4}) which yields a non-zero solution for the second component of $\boldsymbol{\mathcal{S}}$. Since, $\delta\,\sigma_{\pm}^{\,(B)}\,=\,\sigma_{\pm}^{\,(B)}$, as shown in eq. (\ref{e_5.d.17}), the equation concerned reads, 
\begin{align}
&\dot{\big<\sigma_{-}^{(B)}(\tau^{\prime})\,\sigma_{+}^{(B)}(\tau+\tau^{\prime})\big>}\,=\,M_{22}\,\big<\,\sigma_{-}^{(B)}(\tau^{\prime})\,\sigma_{+}^{(B)}(\tau+\tau^{\prime})\big>~;\label{e.6.a.7}
\end{align}
which comes up with the following correlation function
\begin{align}
\big<\sigma_{-}^{\,(B)}&(\tau^{\prime})\,\sigma_{+}^{\,(B)}(\tau+\tau^{\prime})\big>\nonumber\\
&=\big<\sigma_{-}^{\,(B)}(\tau^{\prime})\,\sigma_{+}^{\,(B)}(\tau^{\prime})\big>\,e^{M_{22}\tau}\nonumber\\
&=\frac{\mathcal{G}(\Omega)}{\mathcal{G}_{+}(\Omega)}\Big[\,1-\,e^{-\frac{\mu^2}{4}\mathcal{G}_{+}(\Omega)\tau^{\prime}} \Big]\nonumber\\
&\times\,e^{\big[-\frac{\mu^2}{2}\,\big(\mathcal{G}(0)+\frac{1}{4}\mathcal{G}_{+}(\Omega) \big)+\,i\,\big(2\,\Omega+\frac{\mu^2}{8}K_{-}(\Omega) \big)\big]\tau}~.\label{e.6.a.8}
\end{align}
The above correlation measures the time correlation of absorption in the system. That means the probability amplitude that it can be re-excited at a later time $\tau^{\prime}+\tau$ after having decayed at an earlier time $\tau^{\prime}$. The probability amplitude, real part of the above expression, is increasing with respect to $\tau^{\prime}$ and accompanied by a coherent phase factor, function of delay time $\tau$. That means if an absorption occurs at $\tau^{\prime}$, it evolves coherently until $\tau^{\prime}+\tau$. In a two-point correlation, coherent phase evolution basically indicates the phase evolution of a conditional quantum amplitude (which varies exponentially) between two time-ordered events. Although the single time coherence vanishes \big($\big<\sigma_{\pm}^{(B)}(\tau)\big>=0$\big), this coherence can persist since the phase evolution is conditioned on the earlier operation at $\tau^{\prime}$.\\ 
\noindent Here, once again, we intend to refer to eqs. (\ref{e_4.1.9}, \ref{e_4.1.10d} ) to remember that, for a finite, positive and fixed value of $\Omega$, the rate of decay with respect to $\tau$ (time delay), increases with the increasing value of uniform acceleration $a$.\\
\noindent Observing the above expression, it can be immediately recognized that, at the initial time ($\tau^{\prime}=0$), the above correlation does not begin to form since 
\begin{align}
   \big<\sigma_{-}^{\,(B)}(0)\,\sigma_{+}^{\,(B)}(\tau)\big>\,=\,0~.\label{e.6.a.9}
\end{align}
However, for $\Omega>0$, at the stationary state ($\tau^{\prime}\,\rightarrow\,\infty$), it is reduced to 
\begin{align}
\lim_{\tau^{\prime}\rightarrow\infty}\big<&\sigma_{-}^{\,(B)}(\tau^{\prime})\,\sigma_{+}^{\,(B)}(\tau+\tau^{\prime})\big> \nonumber\\
=&\big<\sigma_{-}^{\,(B)}(\tau^{\prime})\,\sigma_{+}^{\,(B)}(\tau+\tau^{\prime})\big>_{ss}\nonumber\\
=& \frac{\mathcal{G}(\Omega)}{\mathcal{G}_{+}(\Omega)}e^{\big[-\frac{\mu^2}{2}\,\big(\mathcal{G}(0)+\frac{1}{4}\mathcal{G}_{+}(\Omega) \big)+\,i\,\big(2\,\Omega+\frac{\mu^2}{8}K_{-}(\Omega) \big)\big]\tau}~.\label{e.6.a.10}
\end{align}
\noindent It is quite clear from the above expression that the coherent phase at stationary state evolves with an amplitude decaying exponentially with respect to delay time $\tau$. Nevertheless, that steady state expression, in zero delay ($\tau=0$) and in long delay ($\tau\rightarrow\infty$), respectively, yields  
\begin{align}
  \big<\sigma_{-}^{\,(B)}(\tau^{\prime})\,&\sigma_{+}^{\,(B)}(\tau^{\prime})\big>_{ss}\\
  &=\,\frac{\mathcal{G}(\Omega)}{\mathcal{G}_{+}(\Omega)}\,=\dfrac{1}{1+e^{-2\pi\Omega/a}}\,=\,P_{1}(\Omega)~;
\end{align}
and
\begin{align}
  \lim_{\tau\rightarrow\infty}\big<\sigma_{-}^{\,(B)}(\tau^{\prime})\,\sigma_{+}^{\,(B)}(\tau+\tau^{\prime})\big>_{ss}\,=\,0~.
\end{align}
Here, it should be mentioned that the above non-zero fraction decreases with the raising acceleration of the battery, as also discussed below eq. (\ref{e_4.1.10d}) and shown in Figure. (\ref{p1}).\\
\noindent The other possible correlation functions for eq. (\ref{e.6.a.4}) are found to be 
\begin{align}
    &\big<\sigma_{-}^{\,(B)}(\tau^{\prime})\,\sigma_{-}^{\,(B)}(\tau+\tau^{\prime})\big>\,=\,\big<(\sigma_{-}^{\,(B)}(\tau^{\prime}))^2\big>\,e^{M_{11}\,\tau}\,=\,0~;\label{e.6.a.11}
\end{align}
and
\begin{align}
    &\,\,\Big<\sigma_{-}^{\,(B)}(\tau')\,\sigma_{+}^{\,(B)}(\tau+\tau')\,\sigma_{-}^{\,(B)}(\tau+\tau^{\prime})\Big>\nonumber\\
    &=\big<\sigma_{-}^{\,(B)}(\tau^{\prime})\big>\dfrac{\mathcal{G}(-\Omega)}{\mathcal{G}_{+}(\Omega)}+\bigg[\bigg(1-\dfrac{\mathcal{G}(-\Omega)}{\mathcal{G}_{+}(\Omega)}\bigg)\big<\sigma_{-}^{\,(B)}(\tau^{\prime})\big>\nonumber\\
    &-\Big<(\sigma_{-}^{\,(B)}(\tau^{\prime}))^2\sigma_{+}^{\,(B)}(\tau^{\prime})\Big>\bigg]e^{\,M_{33}\tau}\nonumber\\
    &=0~.\label{e.6.a.12}
\end{align}
In a similar way, we can also treat eq. (\ref{e.6.a.5}) which yields the non-zero two time average for the first component of $\boldsymbol{\mathcal{S}}$. That is given by 
\begin{align}
&\,\,\big<\sigma_{+}^{\,(B)}(\tau^{'})\,\sigma_{-}^{\,(B)}(\tau+\tau^{'})\big>
=\big<\sigma_{+}^{\,(B)}(\tau^{'})\,\sigma_{-}^{\,(B)}(\tau^{'})\big>\,e^{M_{11}\tau}\nonumber\\
&=\dfrac{1}{\mathcal{G}_{+}(\Omega)}\Big[\mathcal{G}(-\Omega)+\mathcal{G}(\Omega)\,e^{-\frac{\mu^2}{4}\mathcal{G}_{+}(\Omega)\tau^{\prime}} \Big]\nonumber\\
&\times\,\e^{-\Big[\frac{\mu^2}{2}\,\Big(\mathcal{G}(0)+\frac{\mathcal{G}_{+}(\Omega)}{4} \Big)+\,i\,\Big(2\,\Omega+\frac{\mu^2}{8}K_{-}(\Omega) \Big)\Big]\tau}~.\label{e.6.a.13}
\end{align}
The above correlation measures the time correlation of emission in the system. Once again, for $\Omega>0$, it appears as a combination of a coherent phase factor and an exponentially decaying (with respect to $\tau^{\prime}$) probability amplitude for spontaneous emission at a later time $\tau^{\prime}+\tau$ after having excited at an earlier time $\tau^{\prime}$. Apart from it, as discussed in the previous case (eq. (\ref{e.6.a.8})) also, the decay rate in the above expression increases with the acceleration of the battery for a finite, positive and fixed value of $\Omega$.\\   
\noindent The first notable distinction from the previous one (eq. (\ref{e.6.a.8})) is that the above correlation still exists at the initial time ($\tau^{\prime}=0$) since 
\begin{align}
   \big<\sigma_{+}^{\,(B)}(0)\sigma_{-}^{\,(B)}(\tau)&\big>\nonumber\\
   =\,&e^{-\Big[\frac{\mu^2}{2}\,\Big(\mathcal{G}(0)+\frac{\mathcal{G}_{+}(\Omega)}{4} \Big)+\,i\,\Big(2\,\Omega+\frac{\mu^2}{8}K_{-}(\Omega) \Big)\Big]\tau}~.
\end{align}
This distinction appears consistently with the chosen initial condition that the probability of finding the system in the excited state is $1$. Hence, it contributes in calculating the correlation  $\big<\sigma_{+}^{\,(B)}(0)\sigma_{-}^{\,(B)}(\tau)\big>$ where excitation occurs at $\tau^{\prime}=0$. In contrast, the earlier operation was de-excitation at initial time in the correlation $\big<\sigma_{-}^{\,(B)}(0)\sigma_{+}^{\,(B)}(\tau)\big>$. Since, the probability of de-excitation at $\tau^{\prime}=0$ is zero, it results in zero in eq. (\ref{e.6.a.9}).\\ 
\noindent However, after approaching the steady state ($\tau^{\prime}\,\rightarrow\,\infty$), for $\Omega>0$, it is reduced to be 
\begin{align}
&\lim_{\tau^{\prime}\rightarrow\infty}\big<\sigma_{+}^{\,(B)}(\tau^{\prime})\,\sigma_{-}^{\,(B)}(\tau+\tau^{\prime})\big> \nonumber\\
=&\big<\sigma_{+}^{\,(B)}(\tau^{\prime})\,\sigma_{-}^{\,(B)}(\tau+\tau^{\prime})\big>_{ss}\nonumber\\
=& \frac{\mathcal{G}(-\Omega)}{\mathcal{G}_{+}(\Omega)}e^{-\big[\frac{\mu^2}{2}\,\big(\mathcal{G}(0)+\frac{1}{4}\mathcal{G}_{+}(\Omega) \big)+\,i\,\big(2\,\Omega+\frac{\mu^2}{8}K_{-}(\Omega) \big)\big]\tau}~;
\end{align}
which, in zero delay ($\tau=0$) and in long delay ($\tau\rightarrow\infty$), respectively, yields   
\begin{align}
  \big<\sigma_{+}^{\,(B)}(\tau^{\prime})\,&\sigma_{-}^{\,(B)}(\tau^{\prime})\big>_{ss}\,\nonumber\\
  &=\frac{\mathcal{G}(-\Omega)}{\mathcal{G}_{+}(\Omega)}\,=\,\dfrac{1}{1+e^{2\pi\Omega/a}}\,=\,P_2(\Omega)~,
  \end{align}
and 
\begin{align}
  \lim_{\tau\rightarrow\infty}\big<\sigma_{+}^{\,(B)}(\tau^{\prime})\,\sigma_{-}^{\,(B)}(\tau+\tau^{\prime})\big>_{ss}\,=\,0~.
\end{align}
Once again, we intend to remember that the above non-zero fraction increases with the raising acceleration of the battery, as also discussed below eq. (\ref{e_4.1.10d}) and shown in Figure. (\ref{p2}).\\
\noindent It should be noted that the other possible correlation functions corresponding to eq. (\ref{e.6.a.5}) are zero. 
\begin{align}
    &\big<\sigma_{+}^{\,(B)}(\tau^{\prime})\,\sigma_{+}^{\,(B)}(\tau+\tau^{\prime})\big>\,=\,\big<(\sigma_{+}^{\,(B)}(\tau^{\prime}))^2\big>\,e^{M_{22}\,\tau}\,=\,0~;\label{e.6.a.15} 
\end{align}
and
\begin{align}
    &\,\,\Big<\sigma_{+}^{\,(B)}(\tau^{\prime})\,\sigma_{+}^{\,(B)}(\tau+\tau')\,\sigma_{-}^{\,(B)}(\tau+\tau')\Big>\nonumber\\
    &=\big<\sigma_{+}^{\,(B)}(\tau^{\prime})\big>\dfrac{\mathcal{G}(-\Omega)}{\mathcal{G}_{+}(\Omega)}-\bigg[\dfrac{\mathcal{G}(-\Omega)}{\mathcal{G}_{+}(\Omega)}\big<\sigma_{+}^{\,(B)}(\tau^{\prime})\big>\nonumber\\
    &-\Big<(\sigma_{+}^{\,(B)}(\tau^{\prime}))^2\sigma_{-}^{\,(B)}(\tau^{\prime})\Big>\bigg]e^{\,M_{33}\tau}\nonumber\\
    &=0~.\label{e.6.a.16}
\end{align}
Finally, we attempt to find the last non-zero correlation function of first order from eq. (\ref{e.6.a.10}) which offers a non-zero solution for the third component of 
$\boldsymbol{\mathcal{S}}$. Therefore,
\begin{align}
\Big<\delta\sum(\tau^{\prime})\,\delta\sum(\tau+\tau^{\prime})\Big>
=\Big<\delta\sum(\tau^{\prime})\,\delta\sum(\tau^{\prime})\Big>\,e^{M_{33}\tau}~.\label{e.6.a.17}
\end{align}
Substituting eq. (\ref{e_5.d.18}) in the above solution and using $\big(\sigma_{+}^{\,(B)}\sigma_{-}^{\,(B)}\big)^2\,=\,\sigma_{+}^{\,(B)}\sigma_{-}^{\,(B)}$ for the purpose of simplification, we have 
\begin{align}
&\Big<\sum(\tau^{\prime})\,\sum(\tau+\tau^{\prime})\Big>\nonumber\\
&=\Big<\sigma_{+}^{\,(B)}(\tau^{\prime})\,\sigma_{-}^{\,(B)}(\tau^{\prime})\,\sigma_{+}^{\,(B)}(\tau+\tau^{\prime})\,\sigma_{-}^{\,(B)}(\tau+\tau^{\prime})  \Big>\nonumber\\
&=\dfrac{\mathcal{G}(\Omega)\,\mathcal{G}(-\Omega)}{(\mathcal{G}_{+}(\Omega))^2}\,\Bigg[ e^{-\dfrac{\mu^2}{4}\mathcal{G}_{+}(\Omega)\,\tau}+e^{-\dfrac{\mu^2}{4}\mathcal{G}_{+}(\Omega)\,\tau^{\prime}}\Bigg]\nonumber\\
&\,+\Bigg(\dfrac{\mathcal{G}(\Omega)}{\mathcal{G}_{+}(\Omega)}\Bigg)^2\,e^{-\dfrac{\mu^2}{4}\mathcal{G}_{+}(\Omega)\,(\tau^{\prime}+\tau)}+\Bigg(\dfrac{\mathcal{G}(-\Omega)}{\mathcal{G}_{+}(\Omega)}\Bigg)^2~.\label{e.6.a.18} 
\end{align}
The above correlation basically exists between two measurement of the quantity of excitation stored (probability of existing in excited state) at different points on time axis, namely the earlier time $\tau^{\prime}$ and the later time $\tau^{\prime}+\tau$. Physically, it represents the memory of excitation which measures how strongly the system being in the excited state at time $\tau^{\prime}$ is correlated with it being excited again at time $\tau^{\prime}+\tau$. Although, unlike the previous cases, the probability amplitude does not exhibit any phase evolution, the decay rate varies exponentially (according to eq. (\ref{e_4.1.10d})) in proportion to the acceleration of the battery in a manner similar to the previous scenarios. \\
\noindent One of the important characteristics of the above correlation function is that it can never be reduced to zero for a positive and finite value of $\Omega$. However, $\tau^{\prime}=0$, it just reduces to $\sum(\tau)$. Thus,  
\begin{align}
\Big<\sigma_{+}^{\,(B)}(0)\,\sigma_{-}^{\,(B)}(0)\,\sigma_{+}^{\,(B)}(\tau)\,\sigma_{-}^{\,(B)}(\tau)\Big>=\Big<\sigma_{+}^{\,(B)}(\tau)\,\sigma_{-}^{\,(B)}(\tau)\Big>~;
\end{align}
which, in zero delay ($\tau=0$) and in long delay ($\tau\rightarrow\infty$), respectively, yields   
\begin{align}
\Big<\sigma_{+}^{\,(B)}(0)\,\sigma_{-}^{\,(B)}(0)\,\sigma_{+}^{\,(B)}(0)\,&\sigma_{-}^{\,(B)}(0)\Big>\nonumber\\
=&\Big<\sigma_{+}^{\,(B)}(0)\,\sigma_{-}^{\,(B)}(0)\Big>\,=\,1~,
\end{align}
and
\begin{align}
\lim_{\tau\rightarrow\infty}\Big<\sigma_{+}^{\,(B)}(0)\,&\sigma_{-}^{\,(B)}(0)\,\sigma_{+}^{\,(B)}(\tau)\,\sigma_{-}^{\,(B)}(\tau)\Big>\nonumber\\
=&\lim_{\tau\rightarrow\infty}\Big<\sigma_{+}^{\,(B)}(\tau)\,\sigma_{-}^{\,(B)}(\tau)\Big>\nonumber\\
&=\dfrac{\mathcal{G}(-\Omega)}{\mathcal{G}_{+}(\Omega)}\,=\,\dfrac{1}{1+e^{2\pi\Omega/a}}\,=\,P_2(\Omega)~.
\end{align}
In the stationary state, ($\tau^{\prime}\,\rightarrow\,\infty$), for $\Omega>0$, eq. (\ref{e.6.a.18}) takes the form 
\begin{align}
&\Big<\sum(\tau^{\prime})\,\sum(\tau+\tau^{\prime})\Big>_{ss}=\dfrac{\mathcal{G}(-\Omega)}{\mathcal{G}_{+}(\Omega)}\,\big<\sigma_{+}^{\,(B)}(\tau)\sigma_{-}^{\,(B)}(\tau) \big>~;
\end{align}
which, in zero delay ($\tau=0$) and in long delay ($\tau\rightarrow\infty$), respectively, yields   
\begin{align}
\Big<\sum(\tau^{\prime})\,\sum(\tau^{\prime})\Big>_{ss}=\dfrac{\mathcal{G}(-\Omega)}{\mathcal{G}_{+}(\Omega)}\,&=\,\dfrac{1}{1+e^{2\pi\Omega/a}}\,=\,P_2(\Omega),
\end{align}
and
\begin{align}
&\lim_{\tau\rightarrow\,\infty}\Big<\sum(\tau^{\prime})\,\sum(\tau^{\prime}+\tau)\Big>_{ss}\nonumber\\
&=\bigg(\dfrac{\mathcal{G}(-\Omega)}{\mathcal{G}_{+}(\Omega)}\bigg)^2=\,\dfrac{1}{\big(1+e^{2\pi\Omega/a}\big)^2}\,=\,\big(P_2(\Omega)\big)^2~. 
\end{align}
Here, we should again recall that for a positive and finite value of $\Omega$, $\mathcal{G}(-\Omega)\big/\mathcal{G}_{+}(\Omega)$ is a positive fraction \big(eq. (\ref{e_4.1.10b})\big). As also discussed below eq. (\ref{e_4.1.10b}), the above fraction increases as we increase the acceleration of the battery.\\
\noindent Nevertheless, the other possible correlations associated with eq. (\ref{e.6.a.10}) correspond to the first and second components of $\boldsymbol{\mathcal{S}}$. Those are obtained as 
\begin{align}
    &\big<\sigma_{+}^{\,(B)}(\tau^{\prime})\,\sigma_{-}^{\,(B)}(\tau^{\prime})\,\sigma_{-}^{\,(B)}(\tau+\tau^{\prime})\big>\nonumber\\
    =&\big<\sigma_{-}^{\,(B)}(\tau+\tau^{\prime})\big>\dfrac{\mathcal{G}(-\Omega)}{\mathcal{G}_{+}(\Omega)}+\bigg[-\dfrac{\mathcal{G}(-\Omega)}{\mathcal{G}_{+}(\Omega)}\big<\sigma_{-}^{\,(B)}(\tau^{\prime})\big>\nonumber\\
    &+\Big<\sigma_{+}^{\,(B)}(\tau^{\prime})(\sigma_{-}^{\,(B)}(\tau^{\prime}))^2\Big>\bigg]e^{\,M_{33}\tau} \nonumber\\
    &=\,0~;
\end{align} 
and 
\begin{align}
    &\big<\sigma_{+}^{\,(B)}(\tau^{\prime})\,\sigma_{-}^{\,(B)}(\tau^{\prime})\,\sigma_{+}^{\,(B)}(\tau+\tau^{\prime})\big>\,\nonumber\\
    =&\big<\sigma_{+}^{\,(B)}(\tau+\tau^{\prime})\big>\dfrac{\mathcal{G}(-\Omega)}{\mathcal{G}_{+}(\Omega)}+\bigg[\Big(1-\dfrac{\mathcal{G}(-\Omega)}{\mathcal{G}_{+}(\Omega)}\Big)\big<\sigma_{+}^{\,(B)}(\tau^{\prime})\big>\nonumber\\
    &-\Big<(\sigma_{+}^{\,(B)}(\tau^{\prime}))^2\sigma_{-}^{\,(B)}(\tau^{\prime})\Big>\bigg]e^{\,M_{33}\tau} \nonumber\\
    &=\,0~.
\end{align}
\noindent In an effort to verify the consistency in relation to $\big\lbrace\sigma_{+}^{\,(B)}, \sigma_{-}^{\,(B)}\big\rbrace\,=\,1$, we can also deduce the following correlation,
\begin{align}
\Big<\sum(\tau^{\prime})\,\Delta&(\tau+\tau^{\prime})\Big>\nonumber\\
=\Big<\sigma_{+}^{\,(B)}(\tau^{\prime})\,&\sigma_{-}^{\,(B)}(\tau^{\prime})\,\sigma_{-}^{\,(B)}(\tau+\tau^{\prime})\,\sigma_{+}^{\,(B)}(\tau+\tau^{\prime})  \Big>\nonumber\\
=\frac{\mathcal{G}(\Omega)}{(\mathcal{G}_{+}(\Omega))^2}\,&\Bigg[\mathcal{G}(\Omega) e^{-\frac{\mu^2}{4}\mathcal{G}_{+}(\Omega)\,\tau^{\prime}}+\mathcal{G}(-\Omega) \Bigg]\nonumber\\
&\times\bigg(1-e^{-\frac{\mu^2}{4}\mathcal{G}_{+}(\Omega)\,\tau} \bigg)~.  \label{e.6.a.22}
\end{align}
First, the summation of above two correlation functions \big(eq. (\ref{e.6.a.18}) and eq. (\ref{e.6.a.22})\big) shows   
\begin{align}
&\,\,\Big<\sum(\tau^{\prime})\,\sum(\tau+\tau^{\prime})\Big>+\Big<\sum(\tau^{\prime})\,\Delta(\tau+\tau^{\prime})\Big>\nonumber\\
&=\frac{\mathcal{G}(\Omega)\mathcal{G}(-\Omega)}{(\mathcal{G}_{+}(\Omega))^2}\,e^{-\frac{\mu^2}{4}\mathcal{G}_{+}(\Omega)\,\tau^{\prime}}+\frac{\mathcal{G}(\Omega)\mathcal{G}(-\Omega)}{(\mathcal{G}_{+}(\Omega))^2}\nonumber\\
&+\Bigg(\frac{\mathcal{G}(-\Omega)}{\mathcal{G}_{+}(\Omega)}\Bigg)^2+\Bigg(\frac{\mathcal{G}(\Omega)}{\mathcal{G}_{+}(\Omega)}\Bigg)^2\,e^{-\frac{\mu^2}{4}\mathcal{G}_{+}(\Omega)\,\tau^{\prime}}~.
\end{align}
Next, from the relation $\mathcal{G}_{+}(\Omega)\,=\,\mathcal{G}(\Omega)+\mathcal{G}(-\Omega)$, substituting $\mathcal{G}(-\Omega)$ from the first term and $\mathcal{G}(\Omega)$ from the second term, the above expression is simplified to be
\begin{align}
&\Big<\sum(\tau^{\prime})\,\sum(\tau+\tau^{\prime})\Big>+\Big<\sum(\tau^{\prime})\,\Delta(\tau+\tau^{\prime})\Big>\nonumber\\
&=\frac{1}{\mathcal{G}_{+}(\Omega)}\Big[\mathcal{G}(-\Omega)+\mathcal{G}(\Omega)\,e^{-\frac{\mu^2}{4}\mathcal{G}_{+}(\Omega)\tau^{\prime}} \Big]\nonumber\\
&=\Big<\sum(\tau^{\prime})\Big>\nonumber\\
&=\Big<\sigma_{+}^{\,(B)}(\tau^{\prime})\,\sigma_{-}^{\,(B)}(\tau^{\prime})\Big>~.
\end{align}
The most important physical quantity that can be measured from one of those above first order correlation functions is the power spectrum, which will be discussed in a later section.
\subsection{Second-order correlation function: HBT effect}\label{S:6.3} 
\begin{figure}[h!]
\begin{center}
\includegraphics[scale=0.66]{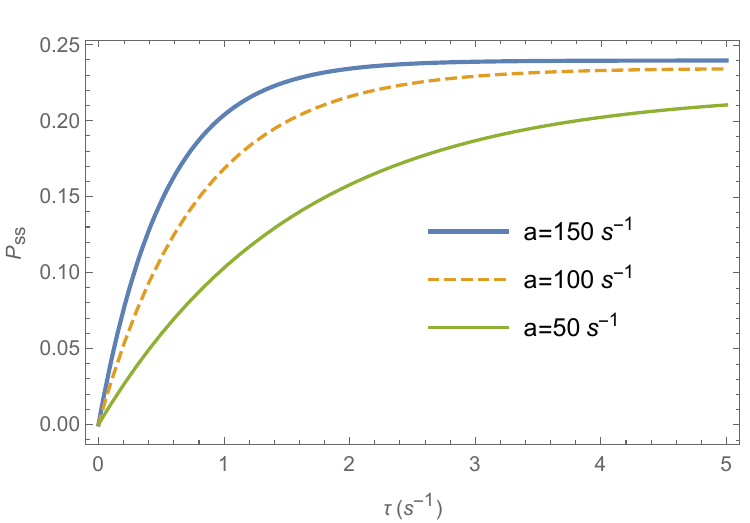}
\caption{Plot of the two point steady-state correlation function of second order (associated with HBT effect) as we vary time delay $\tau$ for different numerical values of uniform acceleration $a$ and for $\Omega=1, \mu\,=1$. \label{pss}}
\end{center}
\end{figure}

\noindent We begin by recalling the general equation of motion of the two time average of three operators evolved in two different points on the time axis. It reads ($\tau\,>0$) 
\begin{align} 
\dfrac{d}{d\tau}\Big<\mathcal{O}_{1}(\tau^{\prime})\,\delta\,\boldsymbol{\mathcal{S}}(\tau+&\tau^{\prime})\mathcal{O}_{2}(\tau^{\prime})\Big>\nonumber\\=&M\Big<\mathcal{O}_{1}(\tau^{\prime})\,\delta\,\boldsymbol{\mathcal{S}}(\tau+\tau^{'})\mathcal{O}_{2}(\tau^{\prime})\Big>~;\label{e.6.b.1} 
\end{align}
where $\mathcal{O}_{1}$ and $\mathcal{O}_{2}$ are any two operators acting in the Hilbert space of the system.\\
\noindent In providing the example of the second-order correlation function, we choose a specific set of operators. For the third component of $\boldsymbol{\mathcal{S}}$, we fix $\mathcal{O}_{1}\,=\,\sigma_{+}^{\,(B)}$ and $\mathcal{O}_{2}\,=\,\sigma_{-}^{\,(B)}$. Here, we should mention that this combination is relevant in the context of the \textit{photon bunching} or the \textit{Hanbury-Brown-Twiss (HBT)} \cite{carmichael1999statistical} effect, a pivotal phenomenon in quantum optics. Our aim is to compare the outcome for our two-level system, which essentially mimics fermionic statistics, with the existing results in the literature for a multi-level system that mimics bosonic statistics.\\[2pt]
\noindent Therefore, the correlation function concerned is to be derived from the following equation.
\begin{align}
\frac{d}{d\tau}\Big<\sigma_{+}^{\,(B)}(\tau^{\prime})\,&\delta\,\sum(\tau+\tau^{\prime})\,\sigma_{-}^{\,(B)}(\tau^{\prime})\Big>\nonumber\\=&\,M_{33}\,\Big<\sigma_{+}^{\,(B)}(\tau^{\prime})\,\delta\,\sum(\tau+\tau^{\prime})\,\sigma_{-}^{\,(B)}(\tau^{\prime})\Big>~;\label{e.6.b.2}
\end{align}
which immediately yields
\begin{align}
&\,\,\Big<\sigma_{+}^{\,(B)}(\tau^{\prime})\,\delta\,\sum(\tau+\tau^{\prime})\,\sigma_{-}^{\,(B)}(\tau^{\prime})\Big>\nonumber\\
&=\,e^{-\frac{\mu^2}{4}\mathcal{G}_{+}(\Omega)\,\tau}\Big<\sigma_{+}^{\,(B)}(\tau^{\prime})\,\delta\,\sum(\tau^{\prime})\,\sigma_{-}^{\,(B)}(\tau^{\prime})\Big>~.\label{e.6.b.3}
\end{align}
Using eq. (\ref{e_5.d.18}) and following the procedure described in the previous discussion, we obtain  
\begin{align}
\,\,\Big<\sigma_{+}^{\,(B)}(\tau^{\prime})\,&\sum(\tau+\tau^{\prime})\,\sigma_{-}^{\,(B)}(\tau^{\prime})\Big>\nonumber\\
=\Big<\sigma_{+}^{\,(B)}(\tau^{\prime})\,&\sigma_{+}^{\,(B)}(\tau+\tau^{\prime})\,\sigma_{-}^{\,(B)}(\tau+\tau^{\prime})\,\sigma_{-}^{\,(B)}(\tau^{\prime})\Big>\nonumber\\
=\frac{\mathcal{G}(-\Omega)}{\big(\mathcal{G}_{+}(\Omega)\big)^2}\,&\bigg(1-e^{-\frac{\mu^2}{4}\mathcal{G}_{+}(\Omega)\,\tau}\bigg) \nonumber\\
&\times \bigg[\mathcal{G}(-\Omega)+\mathcal{G}(\Omega)\,e^{-\frac{\mu^2}{4}\mathcal{G}_{+}(\Omega)\,\tau^{\prime}} \bigg]~.\label{e.6.b.4}
\end{align}
At the initial time ($\tau^{\prime}=0$), the above correlation is reduced to be 
\begin{align}
   \Big<\sigma_{+}^{\,(B)}(0)\,\sigma_{+}^{\,(B)}(\tau)\,&\sigma_{-}^{\,(B)}(\tau)\,\sigma_{-}^{\,(B)}(0)\Big>\nonumber\\
   =&\frac{\mathcal{G}(-\Omega)}{\mathcal{G}_{+}(\Omega)}\,\bigg(1-e^{-\frac{\mu^2}{4}\mathcal{G}_{+}(\Omega)\,\tau}\bigg)~;\label{e.6.b.5}
\end{align}
which, for $\Omega\,>0$, in zero delay ($\tau=0$) and in long delay ($\tau\rightarrow\infty$), respectively, yields   
\begin{align}
\Big<\sigma_{+}^{\,(B)}(0)\,\sigma_{+}^{\,(B)}(0)\,\sigma_{-}^{\,(B)}(0)\,\sigma_{-}^{\,(B)}(0)\Big>=0~,\label{e.6.b.5a}
\end{align}
and
\begin{align}
   \lim_{\tau\rightarrow\infty}\Big<\sigma_{+}^{\,(B)}(0)\,&\sigma_{+}^{\,(B)}(\tau)\,\sigma_{-}^{\,(B)}(\tau)\,\sigma_{-}^{\,(B)}(0)\Big>\nonumber\\
   &=\frac{\mathcal{G}(-\Omega)}{\mathcal{G}_{+}(\Omega)}=\frac{1}{1+\exp\big[\frac{2\pi\Omega}{a}\big]}\,=P_{2}(\Omega)~.\label{e.6.b.6}
\end{align}
However, at the steady state, ($\tau^{\prime}\,\rightarrow\,\infty$), the correlation in eq. (\ref{e.6.b.4}), for $\Omega\,>0$, is reduced to be 
\begin{align}
P_{SS}\nonumber\\
=&\Big<\sigma_{+}^{\,(B)}(\tau^{\prime})\,\sigma_{+}^{\,(B)}(\tau+\tau^{\prime})\,\sigma_{-}^{\,(B)}(\tau+\tau^{\prime})\,\sigma_{-}^{\,(B)}(\tau^{\prime})\Big>_{ss}\nonumber\\
=&\frac{1}{\Big(1+\exp\big[\frac{2\pi\Omega}{a}\big]\Big)^2}\,\bigg(\,1-e^{-\frac{\mu^2\Omega}{8\pi}\,\coth\left(\frac{\pi\,\Omega} {a}\right)\,\tau}\bigg)~.\label{e.6.b.7}
\end{align}
The graphical representation of eq. (\ref{e.6.b.7}) is shown in Figure \ref{pss}. For various numerical values of the uniform acceleration $a$ (in natural units), we plot the variation of steady-state correlation function with respect to the time delay $\tau$. As evident from the graph, for a particular value of $\tau$, the correlation increases as the value of the uniform acceleration increases.\\
\\
The above expression, for $\Omega\,>0$, in zero delay ($\tau=0$) and in long delay ($\tau\rightarrow\infty$), respectively, yields   
\begin{align}
\Big<\sigma_{+}^{\,(B)}(\tau^{\prime})\,\sigma_{+}^{\,(B)}(\tau^{\prime})\,\sigma_{-}^{\,(B)}(\tau^{\prime})\,\sigma_{-}^{\,(B)}(\tau^{\prime})\Big>_{ss}\,=\,0~,\label{e.6.b.7a}
\end{align}
and
\begin{align}
\lim_{\tau\rightarrow\infty}\Big<\sigma_{+}^{\,(B)}(\tau^{\prime})\,&\sigma_{+}^{\,(B)}(\tau+\tau^{\prime})\,\sigma_{-}^{\,(B)}(\tau+\tau^{\prime})\,\sigma_{-}^{\,(B)}(\tau^{\prime})\Big>_{ss}\nonumber\\
   =\bigg[\frac{\mathcal{G}(-\Omega)}{\mathcal{G}_{+}(\Omega)}\bigg]^2=&\dfrac{1}{\bigg(1\,+\,\exp\big[\frac{2\pi\Omega}{a}\big]\bigg)^2}\,=\big(P_{2}(\Omega)\big)^2\nonumber\\
   \equiv\,&\,P_{HBT}
   \label{e.6.b.8}
\end{align}
\begin{figure}[h!]
\begin{center}
\includegraphics[scale=0.66]{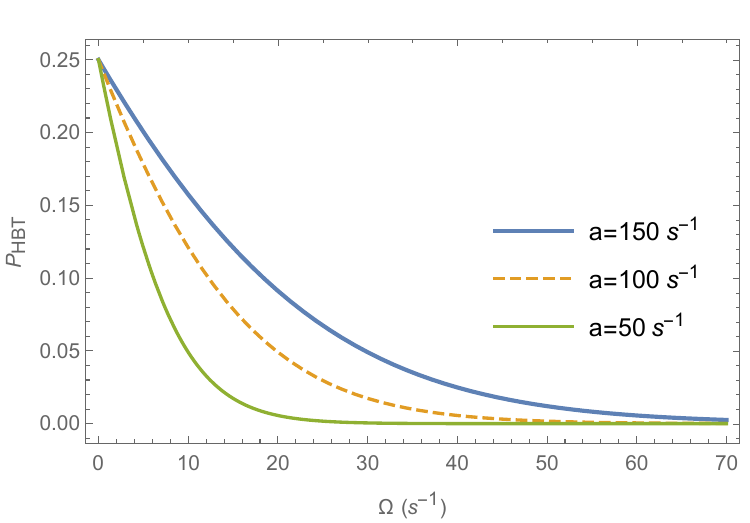}
\caption{Plot of the two point steady-state correlation function of second order (associated with HBT effect) at infinite time delay as we vary frequency $\Omega$ for different numerical values of uniform acceleration $a$. \label{phbt}}
\end{center}
\end{figure}\\
Although Figure \ref{phbt}, which represents the graphical form of the above expression, is a special case of Figure \ref{pss} at $\tau \rightarrow \infty$, and it also increases with increasing uniform acceleration for a fixed value of $\Omega$, the curves exhibit different behaviour compared to those in Figure \ref{pss}.\\[2pt]
\noindent Next, we intend to mention that eq. (\ref{e.6.b.7}) is found to be in contrast with the same found for the bosonic system. As discussed in \cite{carmichael1999statistical}, in the long time, the second order correlation function for a bosonic damped oscillator reads 
\begin{align}
    \lim_{t^{\prime}\rightarrow\infty}\Big<a^{\dagger}(t^{\prime})\,a^{\dagger}(t+t^{\prime})\,a\,(t+t^{\prime})\,a\,(t^{\prime})\Big>\,=\,\bar{n}^2\big(1+e^{\gamma\,t}\big).\label{e.6.b.9}
\end{align}
It provides $2\bar{n}^2$ and $\bar{n}^2$ at zero delay ($t=0$) and long delay ($t\rightarrow\infty$) respectively, where the results take non-zero integer values. In contrast to the plus sign appearing in eq. (\ref{e.6.b.9}), which yields integer outcomes for a multilevel system exhibiting bosonic statistics in time, we find a minus sign in Eq. (\ref{e.6.b.7}), which gives zero and a positive fractional result 
\big(as shown in Eq. (\ref{e_4.1.10b})\big) at zero delay and long delay, respectively. That minus sign also ensures that our result is always less than one, for a finite and positive value of $\Omega$. This captures the typical nature of fermionic statistics and is analogous to the phenomenon of \textit{photon anti-bunching} in the HBT effect. It reflects the fact that a two-level system cannot emit two quanta simultaneously; rather, after one emission, it must be re-excited before emitting again. Hence, the two-level system, which essentially acts as a \textit{single-photon emitter}, shares a deep analogy with the \textit{anti-bunching} phenomenon observed in the HBT effect.\\[2pt]
\noindent Once again, to verify the consistency with respect to $\big\lbrace\sigma_{+}^{\,(B)}, \sigma_{-}^{\,(B)}\big\rbrace\,=\,1$, considering $\boldsymbol{\mathcal{S}}=\Delta=\sigma_{-}^{\,(B)}\sigma_{+}^{\,(B)}$ in eq. (\ref{e.6.b.1}) and using eq. (\ref{e_5.d.21}), we derived
\begin{align}
\Big<\sigma_{+}^{\,(B)}(\tau^{\prime})\,\sigma_{-}^{\,(B)}&(\tau+\tau^{\prime})\,\sigma_{+}^{\,(B)}(\tau+\tau^{\prime})\,\sigma_{-}^{\,(B)}(\tau^{\prime})\Big>\nonumber\\
=\frac{1}{\big(\mathcal{G}_{+}(\Omega)\big)^2}\,&\bigg(\mathcal{G}(-\Omega)+\mathcal{G}(\Omega)e^{-\frac{\mu^2}{4}\mathcal{G}_{+}(\Omega)\,\tau^{\prime}}\bigg) \nonumber\\
&\times \bigg[\mathcal{G}(\Omega)+\mathcal{G}(-\Omega)\,e^{-\frac{\mu^2}{4}\mathcal{G}_{+}(\Omega)\,\tau} \bigg]~.\label{e.6.b.10}
\end{align}
Therefore, adding up both the eq. (\ref{e.6.b.4}) and eq. (\ref{e.6.b.10}), it can be verified that  
\begin{align}
\Big<\sigma_{+}^{\,(B)}(\tau^{\prime})\Big(\sum(\tau+&\tau^{\prime})+\Delta(\tau+\tau^{\prime})\Big)\sigma_{-}^{\,(B)}(\tau^{\prime}) \Big>\nonumber\\
&=\Big<\sum(\tau^{\prime})\Big>=\Big<\sigma_{+}^{\,(B)}(\tau^{\prime})\sigma_{-}^{\,(B)}(\tau^{\prime})\Big>~.\label{e.6.b.11}
\end{align}
\section{Spontaneous emission spectrum}\label{S:7} 


\noindent As a direct quantum analogue of the Wiener–Khinchin theorem \cite{huang, wiener, khintchine, qwkt}, the power spectrum of our battery system can be derived by performing a Fourier transform of the first order correlation function $\Big<\sigma_{+}^{\,(B)}(\tau^{\prime})\,\sigma_{-}^{\,(B)}(\tau+\tau^{\prime})\Big>$. While the correlation function in the time domain describes, after the occurrence of an excitation at $\tau^{\prime}$, the surviving probability of that excitation over a time delay $\tau$, the Fourier transform, in contrast, characterizes the distribution of emission having frequency $\omega$ arising from that excitation.\\
\noindent We first define $(\tau+\tau^{\prime})\,\equiv\,{\tau}^{\prime\prime}$ and rewrite the aforementioned correlation function in the following manner
\begin{align}
\Big<\sigma_{+}^{\,(B)}(\tau^{\prime})\,\sigma_{-}^{\,(B)}(\tau^{\prime\prime})&\Big>\,=\,\frac{\mathcal{G}(-\Omega)}{\mathcal{G}_{+}(\Omega)}\,e^{\big(\,R\,+i\,I\big)\tau^{\prime\prime}}\,e^{-\big(\,R\,+i\,I\big)\tau^{\prime}}\nonumber\\
&+\frac{\mathcal{G}(\Omega)}{\mathcal{G}_{+}(\Omega)}\,e^{\big(\,R\,+i\,I\big)\tau^{\prime\prime}}\,e^{\big(M_{33}-\,R\,-i\,I\big)\tau^{\prime}}~;\label{e.7.1}
\end{align}
where $R$ and $I$ are the real and imaginary parts of $M_{11}$ respectively. Therefore,
\begin{align}
&R=-\frac{\mu^2}{2}\Big(\mathcal{G}(0)+\frac{\mathcal{G}_{+}(\Omega)}{4}\Big)~,~I=-\Big(2\,\Omega+\frac{\mu^2}{8}K_{-}(\Omega)\Big)\label{e.7.2}\\
&\big( M_{33}-R\big)=\frac{\mu^2}{2}\Big(\mathcal{G}(0)-\frac{\mathcal{G}_{+}(\Omega)}{4}\Big) ~.\label{e.7.3}
\end{align}
Since, the power spectrum is defined as the Fourier transformation of $\Big<\sigma_{+}^{\,(B)}(\tau^{\prime})\,\sigma_{-}^{\,(B)}(\tau^{\prime\prime})\Big>$, carrying out the integration that maps the correlator from proper-time $\tau$ to the frequency $\omega$ gives  
\begin{align}
\mathcal{P}(\omega)\,&\equiv\,\int_0^T\,\Big<\sigma_{+}^{\,(B)}(\tau^{\prime})\,\sigma_{-}^{\,(B)}(\tau^{\prime\prime})\Big>\,e^{i\omega\,\big(\tau^{\prime\prime}-\tau^{\prime}\big)}\,d\tau^{\prime\prime}\,d\tau^{\prime}\nonumber\\
&=\mathcal{P}_{1}(\omega)\,+\,\mathcal{P}_{2}(\omega)~;\label{e.7.4}
\end{align}
where $\mathcal{P}_{1}(\omega)$ has the following structure 
\begin{widetext}
\begin{align}
\mathcal{P}_{1}(\omega)&=\frac{\mathcal{G}(-\Omega)}{\mathcal{G}_{+}(\Omega)}
\times\frac{{\dfrac{\mu^4}{4}\Big(\mathcal{G}(0)+\frac{\mathcal{G}_{+}(\Omega)}{4}\Big)^2-\Big(\omega-\,2\Omega-\frac{\mu^2}{8}K_{-}(\Omega)\Big)^2+i\mu^2\,\Big(\mathcal{G}(0)+\frac{\mathcal{G}_{+}(\Omega)}{4}\Big) \Big(\omega-\,2\Omega-\frac{\mu^2}{8}K_{-}(\Omega)\Big)}}{{\Bigg[ \frac{\mu^4}{4}\Big(\mathcal{G}(0)+\frac{\mathcal{G}_{+}(\Omega)}{4}\Big)^2-\Big(\omega-\,2\Omega-\frac{\mu^2}{8}K_{-}(\Omega)\Big)^2 \Bigg]^2+\mu^4\Big(\mathcal{G}(0)+\frac{\mathcal{G}_{+}(\Omega)}{4}\Big)^2\,\Big(\omega-\,2\Omega-\dfrac{\mu^2}{8}K_{-}(\Omega)\Big)^2}}\nonumber\\\nonumber\\
&\times\,\bigg\{e^{\Big[-\frac{\mu^2}{2}\Big(\mathcal{G}(0)+\frac{\mathcal{G}_{+}(\Omega)}{4}\Big)\,+i\,\Big(\omega-\,2\Omega-\frac{\mu^2}{8}K_{-}(\Omega)\Big)\Big]\,T}-2+e^{\Big[\frac{\mu^2}{2}\Big(\mathcal{G}(0)+\frac{\mathcal{G}_{+}(\Omega)}{4}\Big)\,-i\,\Big(\omega-\,2\Omega-\frac{\mu^2}{8}K_{-}(\Omega)\Big)\Big]\,T}\bigg\}~;\label{e.7.5}
\end{align}
and $\mathcal{P}_{2}(\omega)$ has the following structure
\begin{align}
\mathcal{P}_{2}(\omega)&
=\frac{\mathcal{G}(\Omega)}{\mathcal{G}_{+}(\Omega)}
\times\,\frac{{\frac{\mu^4}{4}\Bigg(\frac{{\big(\mathcal{G}_{+}(\Omega)}\big)^2}{16}-{\big(\mathcal{G}(0)\big)}^2\Bigg)+\Big(\omega-\,2\Omega-\frac{\mu^2}{8}K_{-}(\Omega)\Big)^2-\,i\,\mu^2\,\mathcal{G}(0) \Big(\omega-\,2\Omega-\frac{\mu^2}{8}K_{-}(\Omega)\Big)}}{{\Bigg[ \frac{\mu^4}{4}\bigg(\frac{{\big(\mathcal{G}_{+}(\Omega)}\big)^2}{16}-{\big(\mathcal{G}(0)\big)}^2\bigg)+\Big(\omega-\,2\Omega-\frac{\mu^2}{8}K_{-}(\Omega)\Big)^2  \Bigg]^2 +\,\mu^4\,{\big(\mathcal{G}(0)\big)}^2\,\Big(\omega-\,2\Omega-\frac{\mu^2}{8}K_{-}(\Omega)\Big)^2}}\nonumber\\\nonumber\\
&\times\,\bigg\{1-e^{\Big[-\frac{\mu^2}{2}\Big(\mathcal{G}(0)+\frac{\mathcal{G}_{+}(\Omega)}{4}\Big)\,+i\,\Big(\omega-2\Omega-\frac{\mu^2}{8}K_{-}(\Omega)\Big)\Big]\,T}+e^{-\frac{\mu^2}{4}\mathcal{G}_{+}(\Omega)\,T}- e^{\Big[\frac{\mu^2}{2}\Big(\mathcal{G}(0)-\frac{\mathcal{G}_{+}(\Omega)}{4}\Big)\,-i\,\Big(\omega-\,2\Omega-\frac{\mu^2}{8}K_{-}(\Omega)\Big)\Big]\,T}\bigg\}~.\label{e.7.6}
\end{align}
\end{widetext}
In the long time limit $T\to\,\infty$, with the aim of looking for a well defined spectrum from the above expression of power spectrum, we investigate the system in a very high frequency regime where $\omega\rightarrow\infty$. Hence, imposing the concept of rapid oscillation, we look at the following consequence
\begin{align}
&\lim_{T\to\infty}e^{i\,\Big(\omega-\,2\Omega-\frac{\mu^2}{8}K_{-}(\Omega)\Big)\,T}\nonumber\\
&=\,i\pi\Big(\omega-\,2\Omega-\frac{\mu^2}{8}K_{-}(\Omega)\Big)\delta\big(\omega-\,2\Omega-\frac{\mu^2}{8}K_{-}(\Omega)\big)\nonumber\\
&=0~;\label{e.7.7}
\end{align}
since $\delta\big(\omega-\,2\Omega-\frac{\mu^2}{8}K_{-}(\Omega)\big)=0$ as $\omega\neq\big(2\Omega+\frac{\mu^2}{8}K_{-}(\Omega)\big)$. For detailed calculation, see Appendix \ref{AppA}.\\[2pt]
\noindent Hence, at $T\to\,\infty$, the real part of $\mathcal{P}_{1}(\omega)$ and $\mathcal{P}_{2}(\omega)$ turns out to be
\begin{widetext}
\begin{equation}
\Re{\mathcal{P}_{1}(\omega)}\Big|_{T\rightarrow\,\infty}\propto\,\frac{\frac{\mu^4}{4}\Big(\mathcal{G}(0)+\frac{\mathcal{G}_{+}(\Omega)}{4}\Big)^2-\Big(\omega-\,2\,\Omega-\frac{\mu^2}{8}K_{-}(\Omega)\Big)^2 }{{\bigg[ \frac{\mu^4}{4}\Big(\mathcal{G}(0)+\frac{\mathcal{G}_{+}(\Omega)}{4}\Big)^2-\Big(\omega-\,2\,\Omega-\dfrac{\mu^2}{8}K_{-}(\Omega)\Big)^2 \bigg]^2 +\mu^4\,\Big(\mathcal{G}(0)+\frac{\mathcal{G}_{+}(\Omega)}{4}\Big)^2\,\Big(\omega-\,2\,\Omega-\frac{\mu^2}{8}K_{-}(\Omega)\Big)^2}}~;\label{e.7.8}
\end{equation}
and 
\begin{equation}
\Re{\mathcal{P}_{2}(\omega)}\Big|_{T\rightarrow\,\infty}\propto\,\frac{\frac{\mu^4}{4}\bigg(\frac{{\big(\mathcal{G}_{+}(\Omega)}\big)^2}{16}-{\big(\mathcal{G}(0)\big)}^2\bigg)+\Big(\omega-\,2\Omega-\frac{\mu^2}{8}K_{-}(\Omega)\Big)^2}{{\bigg[ \frac{\mu^4}{4}\Big(\frac{{\big(\mathcal{G}_{+}(\Omega)}\big)^2}{16}-{\big(\mathcal{G}(0)\big)}^2\Big)+\Big(\omega-\,2\Omega-\frac{\mu^2}{8}K_{-}(\Omega)\Big)^2\bigg]^2+\mu^4\,{\big(\mathcal{G}(0)\big)}^2\,\Big(\omega-\,2\Omega-\frac{\mu^2}{8}K_{-}(\Omega)\Big)^2}}~.\label{e.7.9}
\end{equation}
\end{widetext}
As evident from both the expressions of ${\mathcal{P}_{1}(\omega)}$ and ${\mathcal{P}_{2}(\omega)}$, the emission spectrum ${\mathcal{P}(\omega)}$ exhibits a \textit{Lorentzian line shape} \cite{carmichael1999statistical}. Hence, the spontaneous emission has a well defined spectrum in the regime $\omega\rightarrow\infty$.


\section{Conclusions}\label{S:8}
\noindent In this paper, our primary aim was to employ the \textit{quantum regression theorem (QRT)}, a powerful tool in the study of open quantum systems, to analyse the impact of the interaction between a uniformly accelerated \textit{Unruh–DeWitt (UDW) detector}, which absorbs charges from an external classical pulse and is thus viewed as a \textit{relativistic quantum battery}, and the environmental bath of its perceived particles, namely, the quanta of a massless scalar field. For doing so, we first model the UDW detector as a two-level quantum system, and to upgrade that system into a quantum battery, it is linearly coupled to an external, time-varying classical driving pulse. In the weak-coupling limit, the frequency of the applied driving pulse is tuned to resonate with the detector’s energy gap (intrinsic frequency) in order to utilize the \textit{rotating wave approximation} and construct a primordial dynamical form of the Hamiltonian for the battery. Furthermore, by employing a rotating-frame transformation, the time dependence is removed from that Hamiltonian, and a simpler, time-independent form of the Hamiltonian is prepared as the final form of the system Hamiltonian of the Unruh–DeWitt detector battery, which is ready to accelerate uniformly and interact with the surrounding environment governed by the massless scalar field.\\[2pt]
\noindent Since, the battery is accelerating uniformly with respect to an inertial observer, by considering the particle moving in \textit{Rindler space}, we proceed our analysis. Next, to access the system’s dynamics under the \textit{Born–Markov approximation}, we derive the \textit{Lindblad master equation}, or \textit{GKSL master equation}, for the evolution of the reduced density matrix with respect to the proper time of the relativistic battery. Then, to compute the functional forms appearing in the master equation, we evaluate the Wightman functions and apply a Fourier transformation to map those functions into the frequency (dual space of proper time) space. Although, the real part of the resultant function after Fourier transformation is found to have a well-behaved analytical structure, the imaginary part, which also contributes to the Lamb-shift term in the master equation, is found to be a divergent quantity which is later regularised by using exponential regularisation. At this stage, with all functional forms in the master equation, the master equation is ready to  contribute to the further extension of this study, namely the computation of single time expectation values and two-time correlation functions of the operators regarding the battery system.\\[2pt]
\noindent Since the implementation of the quantum regression theorem in our present study, necessarily involves with the ladder operators and the number operators in the system's Hilbert space, we define the set of ladder operators which would consequently construct the number operator also. Based on the fact that those operators are directly related to both the matrix elements of the master equation and the Pauli operators, we proceed in two ways for obtaining the differential equation governed by the expectation values of those operators concerned. Eventually, by solving those equations under the chosen condition that the system was in the excited state at the initial time, we find the single-time average values of the ladder and number operator. While the single-time average of the ladder operators provides a zero value, that of the number operator exhibits exponential decay with respect to proper time. This indicates that, although there is no coherent superposition or coherent radiation between the two states, the system can still be excited or dynamically active and loses energy to the environment by emitting spontaneous radiation. Hence, the probability of being in the excited state, which is the initial state of the system, decays exponentially with time. Interestingly, the rate of this dissipation is exponentially proportional to the uniform acceleration of the detector, with the other parameters held fixed. This occurs because the detector would perceive and interact with more particles as the value of uniform acceleration increases, resulting in greater dissipation. Next, the differential equations governing the single-time averages are compactly expressed in vector form. Since this equation is not homogeneous, which is an essential requirement to apply the quantum regression theorem directly, by defining the fluctuations about the steady state we construct a set of homogeneous differential equations obeyed by these fluctuation averages, which is now suitable for employing the QRT to determine the two-time correlation functions.\\[3pt]
\noindent The implementation of QRT yields the evolution equations for the fluctuation correlations, which can eventually give rise to the correlators of first order and second order, up to which we restrict our present study. First, we find three non-zero correlators of first order.      
While the first two correlators couple two different ladder operators measured at an earlier and later time, respectively, implying the probabilities of absorption and spontaneous emission, the third one involves the number operator acting twice with a finite time delay, signifying how strongly the probability of persisting the excitation at an earlier time is correlated with the same at a later time. \\
\noindent When the initial time is set to be the earlier time and hence the later time presents the time delay itself, the correlation function involved with the absorption (when de-excitation occurs earlier, followed by an excitation) provides zero value. In contrast, the same involved with the spontaneous emission (when excitation occurs earlier, followed by a de-excitation) which varies with respect to time delay, exhibits a exponential decay accompanied by a phase evolution. This is expected because of the fact that the system was in excitation initially. Hence, the probability of de-excitation at initial time is zero, resulting in the zero value for the correlation. However, in steady state, both the correlators are only the function of time delay and display exponentially decaying amplitude with a phase evolution. Moreover, in the steady-state, both the functions vanish at long delay and, at zero delay, yield two distinct positive fractions whose sum is unity. While the fraction arising from the absorption correlator decreases with the  increasing value of the uniform acceleration of detector, the second fraction behaves oppositely. It is noteworthy to mention that, in each of the aforementioned cases, the decay rate is proportional (though not necessarily exactly linear or exponential) to the value of the uniform acceleration of the detector battery.\\
\noindent Unlike the first two correlators, the third one, which involves with the number operator, does not get reduced to zero under any condition. When the earlier time is set to be the initial time, it behaves as the single-time average of number operator, which is a function of time delay. Apart from it, the steady-state expression of the correlator is also a single time average of number operator multiplied by a positive fraction, resulting in the same quantity in the zero delay limit and the square of that in the long delay limit. The value of the fraction increases with the assigned value of the uniform acceleration of detector.\\
\noindent In order to address the two-time average of the second order, we chose a particular combination of the ladder operator and the number operator, which is generally considered to analyse the \textit{photon bunching} or \textit{Hanbury-Brown-Twiss (HBT)} effect in the multi-level systems which mimic the bosonic statistics in nature. We intend to look for any fermionic signature in the result obtained from the aforementioned correlator. Referring to the steady state expression related to the HBT effect for a damped harmonic oscillator interacting with environment, we obtain the corresponding expression for our system and analyse both of them. While the result for the multi-level system yield two different non-zero integers for zero delay and long delay respectively, our result provides a zero value and a positive fraction at zero delay and long delay respectively. Moreover, apart from those outcomes, our result at steady state, is always less than one. This captures the typical nature of fermionic statistics and resembles the phenomenon of \textit{photon anti-bunching} in the HBT effect. Since, a two-level system cannot emit two quanta simultaneously, it acts as a \textit{single-photon emitter}, shares a deep analogy with the \textit{anti-bunching} phenomenon observed in the HBT effect.\\[2pt] 
\noindent Finally, we proceed to obtain the expression of power spectrum, one of the physically significant quantities that can be derived from the two-time correlator of the spontaneously emitted radiation from the accelerating battery. After mapping the emission correlator from proper-time space to the frequency space through a Fourier transformation, we observed that the spectrum can exhibit a \textit{Lorentzian line shape} at long time if the frequency is chosen to be very high. This is one of the main observations in this work.\\[2pt]
\noindent Our overall study concludes that the UDW battery dissipates, or loses its coherence, due to interaction, where the rate of dissipation increases with the value of its uniform acceleration, and the spontaneous emission spectrum at large times becomes well behaved in the high-frequency regime.\\[2pt]

\begin{appendix}
\section{Rapid oscillation in long time}\label{AppA}
\noindent In this Appendix, we will evaluate the following improper definite integral. This will be important in order to explain the result obtained in eq. (\ref{e.7.7}) in section \ref{S:7}.\\[2pt]
The integral is given by
\begin{align}
&\int_0^\infty\,e^{i\,\Big(\omega-\,2\Omega-\frac{\mu^2}{8}K_{-}(\Omega)\Big)\,T}\,dT\nonumber\\
=&\frac{-i}{\Big(\omega-\,2\Omega-\frac{\mu^2}{8}K_{-}(\Omega)\Big)}\,e^{i\,\Big(\omega-\,2\Omega-\frac{\mu^2}{8}K_{-}(\Omega)\Big)T}\Bigg\vert_{0}^{\infty}                   \nonumber\\
=&\dfrac{-i}{\Big(\omega-\,2\Omega-\frac{\mu^2}{8}K_{-}(\Omega)\Big)}  \bigg[\lim_{T\rightarrow\infty}\,e^{i\,\Big(\omega-\,2\Omega-\frac{\mu^2}{8}K_{-}(\Omega)\Big)T}-\,1\,\bigg]\,.\label{app1} 
\end{align}
Again, we can compute the above integration by using the regularisation technique \cite{peskin2018introduction}.\\
We can write the above integral in the following form
\begin{align}
&\int_0^\infty\,e^{i\,\Big(\omega-\,2\Omega-\frac{\mu^2}{8}K_{-}(\Omega)\Big)\,T}\,dT\nonumber\\
=&\lim_{\epsilon\to 0^{+}}\int_0^\infty\,e^{i\,\Big(\omega-\,2\Omega-\frac{\mu^2}{8}K_{-}(\Omega)\Big)\,T}\,e^{-\epsilon\,T}\,dT\nonumber\\
=&\lim_{\epsilon\to 0^{+}}\int_0^\infty\,e^{\Big[i\,\Big(\omega-\,2\Omega-\frac{\mu^2}{8}K_{-}(\Omega)\Big)-\epsilon\Big]\,T}\,dT\nonumber\\
=&\lim_{\epsilon\to 0^{+}}\Bigg\vert \frac{e^{\Big[i\,\Big(\omega-\,2\Omega-\frac{\mu^2}{8}K_{-}(\Omega)\Big)-\epsilon\Big]\,T}}{i\,\Big(\omega-\,2\Omega-\frac{\mu^2}{8}K_{-}(\Omega)\Big)-\epsilon}\Bigg\vert_{0}^{\infty}\label{app2}~.
\end{align}
Since $\epsilon>0$, the exponential vanishes at infinity and the above equation becomes
\begin{align}
    &\lim_{\epsilon\to 0^{+}}\Bigg\vert \frac{e^{\Big[i\,\Big(\omega-\,2\Omega-\frac{\mu^2}{8}K_{-}(\Omega)\Big)-\epsilon\Big]\,T}}{i\,\big(\omega-\,2\Omega-\frac{\mu^2}{8}K_{-}(\Omega)\big)-\epsilon}\Bigg\vert_{0}^{\infty}\nonumber\\
    =&-i\,\lim_{\epsilon\to 0^{+}}\left[\frac{1}{\big(\omega-\,2\Omega-\frac{\mu^2}{8}K_{-}(\Omega)\big)-i\epsilon} \right]\label{app3}~.
\end{align}
Now we use the \textit{Sokhotski–Plemelj} \cite{sok, plem} identity,
\begin{equation}
    \lim_{\epsilon\to 0^{+}}\frac{1}{x\pm i\epsilon}=\mp i\pi\delta(x)+\mathscr{P}\left(\frac{1}{x}\right)~\label{app4}
\end{equation}
where $\delta(x)$ is the \textit{Dirac delta function} and $\mathscr{P}$ denotes the \textit{Cauchy principal value}.\\[2pt]
Using the above identity, eq. \eqref{app3} takes the form
\begin{align}
    &-i\,\lim_{\epsilon\to 0^{+}}\left[\frac{1}{\big(\omega-\,2\Omega-\frac{\mu^2}{8}K_{-}(\Omega)\big)-i\epsilon} \right]\nonumber\\
    =&\pi\,\delta\Big(\omega-\,2\Omega-\frac{\mu^2}{8}K_{-}(\Omega)\Big)-i\mathscr{P}\bigg(\dfrac{1}{\omega-\,2\Omega-\frac{\mu^2}{8}K_{-}(\Omega)}\bigg)~.\label{app5}
\end{align}
Substituting eq.~\eqref{app5} into eq.~\eqref{app2}, we obtain
\begin{align}
    &\int_0^\infty\,e^{i\,\Big(\omega-\,2\Omega-\frac{\mu^2}{8}K_{-}(\Omega)\Big)\,T}\,dT\nonumber\\
     =&\pi\,\delta\Big(\omega-\,2\Omega-\frac{\mu^2}{8}K_{-}(\Omega)\Big)-i\mathscr{P}\bigg(\dfrac{1}{\omega-\,2\Omega-\frac{\mu^2}{8}K_{-}(\Omega)}\bigg)~.\label{app6}
\end{align}
Comparing the results given in eq.~\eqref{app1} and eq.~\eqref{app6}, we conclude 
\begin{align}
&\lim_{T\rightarrow\infty}\dfrac{e^{i\,\Big(\omega-\,2\Omega-\frac{\mu^2}{8}K_{-}(\Omega)\Big)T}}{i\Big(\omega-\,2\Omega-\frac{\mu^2}{8}K_{-}(\Omega)\Big)}=\pi\delta\Big(\omega-\,2\Omega-\frac{\mu^2}{8}K_{-}(\Omega)\Big)\,;\label{app7}\\
&\mathscr{P}\bigg(\dfrac{1}{\omega-\,2\Omega-\frac{\mu^2}{8}K_{-}(\Omega)}\bigg)=-\dfrac{1}{\Big(\omega-\,2\Omega-\frac{\mu^2}{8}K_{-}(\Omega)\Big)}~.\label{app8}
\end{align}
From eq.~\eqref{app7}, we obtain
\begin{align}
    &\lim_{T\rightarrow\infty}e^{i\,\Big(\omega-\,2\Omega-\frac{\mu^2}{8}K_{-}(\Omega)\Big)T}\nonumber\\
    =&i\pi\Big(\omega-\,2\Omega-\frac{\mu^2}{8}K_{-}(\Omega)\Big)\delta\Big(\omega-\,2\Omega-\frac{\mu^2}{8}K_{-}(\Omega)\Big)~.\label{app9}
\end{align}
Now, in the context of rapid oscillation ($\omega\to\infty$), which suggests $\delta\Big(\omega-\,2\Omega-\frac{\mu^2}{8}K_{-}(\Omega)\Big)=0$ as $\omega\neq \Big(2\Omega-\frac{\mu^2}{8}K_{-}(\Omega)\Big)$. Since, both $\Omega$ and $K_{-}(\Omega)$ are finite quantities, therefore, from eq. \eqref{app9}, we obtain
\begin{align}
&\lim_{T\rightarrow\infty}e^{i\,\Big(\omega-\,2\Omega-\frac{\mu^2}{8}K_{-}(\Omega)\Big)T}\,=\,0~.
\end{align}
\end{appendix}


\bibliographystyle{hephys.bst}
\bibliography{main}

\end{document}